\documentclass[%
 reprint,
 amsmath,amssymb,
 aps,
 onecolumn,
12pt]{revtex4-2}

\usepackage[T1]{fontenc}
\usepackage{mathptmx}
\DeclareMathAlphabet{\mathcal}{OMS}{cmsy}{m}{n} 

\usepackage{multirow}
\usepackage{subcaption}
\usepackage[font=small]{caption}
\usepackage{float}
\usepackage{mathrsfs}
\usepackage{color}
\usepackage{graphicx}
\usepackage{subeqnarray}
\usepackage{dcolumn}
\usepackage{bm}
\usepackage[
scale=0.78, marginratio={1:1, 1:1}, ignoreall,
]{geometry}

\begin{document}

\preprint{APS/123-QED}


\title{\large{Data-driven classification of sheared stratified turbulence \\ from experimental shadowgraphs}}

\author{Adrien Lefauve}
\email{lefauve@damtp.cam.ac.uk}

\author{Miles M. P. Couchman}

\affiliation{Department of Applied Mathematics and Theoretical Physics, University of Cambridge \\ Centre for Mathematical Sciences, Wilberforce Road, Cambridge CB3 0WA, UK} 

\date{4 May 2023}

\begin{abstract}
We propose a dimensionality reduction and unsupervised clustering method for the automatic classification and reduced-order modeling of density-stratified turbulence in laboratory experiments. We apply this method to 113 long shadowgraph movies collected in a `Stratified Inclined Duct' (SID) experiment, where turbulence is generated by instabilities arising from a sheared buoyancy-driven counterflow at Reynolds numbers $Re \approx 300-5000$, tilt angles $\theta=1^\circ-6^\circ$ and Prandtl number $Pr \approx$ 700. The method automatically detects edges representative of discrete density interfaces, extracts a low-dimensional vector of statistics representative of their morphology, projects these statistics onto a two-dimensional phase space of principal coordinates, and applies the OPTICS clustering algorithm. Five clusters are detected and interpreted physically based on their typical interface morphology and an examination of representative frames, revealing distinct types of turbulence and mixing: laminarizing, braided, overturning, granular and unstructured, as well as some intermediate types.  The ratio of time spent in each cluster varies gradually across the $(Re,\theta)$ space. At intermediate values of $Re\,\theta$, intermittent turbulence cycles between clusters in phase space and reveals at least two distinct routes to stratified turbulence. These insights demonstrate the potential of this method to reveal the underlying physics of complex turbulent systems from large experimental datasets. 
\end{abstract}

\maketitle

\vspace{-0.6cm}

\section{Introduction}\label{sec:intro}

\textit{Context and motivation} -- Fluid flows, viewed under the dynamical systems lens, yield many examples of spatially and temporally coherent structures as the strength of nonlinearities within the flow (as quantified by the Reynolds number) increases. Examples include turbulent spots in plane Couette flow, puffs and slugs in pipe flow, bands in channel flow,  vortex streets in the wake of bluff bodies, ring vortices in buoyant plumes, hairpin and horseshoe vortices in boundary layers, and billows and braids in mixing layers \citep{vandyke_album_1982}.  Flows characterised by more than one non-dimensional parameter may exhibit richer dynamical behaviors, with a variety of distinct flow `states' or `regimes' emerging in different regions of their multi-dimensional parameter space. One salient example is the ``surprisingly complex transition diagram'' observed in Taylor-Couette flow, where a plethora of spatio-temporal dynamics are observed in the two-dimensional parameter space described by the Reynolds numbers of the inner and outer rotating cylinders \citep{andereck_flow_1986}. Identifying such regimes and delineating their extent in parameter space is typically performed manually by a trained human eye based on an inspection of various properties of the observed coherent structures. In this Article and companion Letter \cite{lefauve_routes_2023} we demonstrate that the classification of regimes within complex flows, and the physical insight thus gained, can be enhanced by automated data-driven algorithms. We demonstrate this by revealing the rich turbulent states and transitions in an experiment, the `Stratified Inclined Duct', which we argue represents a new paradigm for the study of turbulence, on par with the well-researched cylindrical pipe  \cite{barkley_theoretical_2016} or Taylor-Couette \cite{feldman_routes_2023} experiments.

To understand the potential advantages of our data-driven approach, it is worth summarising how such canonical flows are typically analyzed. Two broad questions are usually considered. First, how can we predict the emergence and evolution of  coherent structures as non-dimensional parameters are varied? This can be done from first principles (e.g. examining bifurcations of the governing equations and stability theory), or through consideration of more approximate and phenomenological models. Second, how do these coherent structures relate to flow phenomena of interest, e.g. wall drag or mixing? These two steps can thus be represented as:  \vspace{0.2cm}

 input  parameters \  $\xrightarrow{\ \ \text{step 1} \ \ }$  \ regimes, coherent structures \  $\xrightarrow{\ \ \text{step 2} \ \ }$ \  useful output variables. \  (1)\\ \vspace{-0.6cm}

\quad  (equations) \hspace*{9.6cm} (solutions) 
\setcounter{equation}{1}
\vspace{0.2cm}

The intermediate focus on regimes and coherent structures is relevant because they generally play a leading-order role in governing the output. Our companion Letter provides a more detailed discussion of the coherent structure approach to modeling turbulence which uses a `skeleton' of `simple invariant solutions' or `exact coherent states'. Uncovering the role of coherent structures allows us to build physical intuition in terms of mechanical processes and cause-and-effect relationships, thus bridging the gap between the governing equations and their solutions. Developing new data-driven techniques to discover distinct regimes and quantify their properties thus seems critical in deepening our understanding of turbulence.

\textit{Focus} -- We focus here on identifying distinct turbulent states in stratified shear flows, that is turbulence energized by a mean shear between two counterflowing fluid layers having slightly different densities (satisfying the Boussinesq approximation). Coupling terms in the equations governing the evolution of the momentum and density fields mean that coherent velocity structures (e.g. shear and vortices), interact with coherent density structures (i.e. sharp density interfaces of enhanced gradient). This interaction is very complex and predictions from first principles are lacking. For instance, vortices may broaden density interfaces and/or actively sharpen them depending on the circumstances \citep{caulfield_layering_2021}. This has leading-order but poorly-understood implications for the energy dissipation, the buoyancy flux across stable density interfaces, irreversible diapycnal mixing and its efficiency, which are variables of central importance in ocean and climate modeling \cite{gregg_mixing_2018}. In other words, progress is needed on steps 1 and 2 in expression (1).

\textit{Choice of problem} -- To improve our understanding of the regimes of sheared stratified turbulence under controlled laboratory conditions (focusing on step 1), we collected data in the `Stratified Inclined Duct' (SID). Insightful experiments on the instability of stratified shear flow have a long history, dating back at least to the seminal papers of Reynolds in 1883 \cite[\S 12]{reynolds_experimental_1883} (who noted ``it proved a very pretty experiment''),  Taylor in 1927 \cite{taylor_experiment_1927}, and Thorpe in 1971 \cite{thorpe_experiments_1971}. The novelty of SID is that it sustains a highly dissipative two-layer exchange flow through a long tilted rectangular duct connecting two large reservoirs of fluid at different densities. It allow us to explore regions of parameter space and record long time series of turbulent dynamics which were previously inaccessible to experiments and which remain prohibitively expensive to simulate numerically. Four regimes have been identified in SID in the two-dimensional space spanned by the Reynolds number and duct tilt angle: stable laminar flow, finite-amplitude interfacial (`Holmboe') waves, intermittent turbulence and full turbulence. These regimes were first described in 1961 \cite{macagno_interfacial_1961}, before being independently rediscovered in 2014 \cite{meyer_stratified_2014}. Since then, measurements of the three-dimensional volumetric velocity and density fields \citep{partridge_versatile_2019} have allowed progress on steps 1 and 2. The regime diagrams were partially explained from first principles using energy budgets and dissipation arguments \citep{lefauve_regime_2019,lefauve_buoyancy_2020,lefauve_experimental2_2022} as well as analytic theory \citep{duran-matute_regime_2023}, and the morphology and interaction of three-dimensional vortices with density interfaces were described and linked to a linear instability \citep{lefauve_structure_2018,jiang_evolution_2022}.

However, a limitation of SID research to date lies in the subjectivity with which the regimes are defined and the flows are classified. This causes at least three problems undermining both steps 1 and 2. First, this classification relies on choosing qualitative visual criteria with which to classify the flow, which are typically inconsistent between different individuals, especially in transitional regions where a flow exhibits elements of multiple regimes. Second, a classification into discrete regimes implicitly implies sharp transitions, whereas a trained eye recognizes that SID transitions smoothly between regimes (e.g. the turbulent periods and their intensity increase across the intermittent regime). Third, assigning a single regime label to the entire temporal evolution of an unsteady flow is reductive and brushes over its spatial and temporal complexity. 

\textit{Approach and outline} -- This paper resolves the problems associated with the subjective nature of a human classification by applying an objective, automated, and physically-interpretable classification frame-by-frame to a large experimental dataset of 113 shadowgraph movies. Recent work has demonstrated that density interfaces and their distribution and structure through a turbulent flow play a key role in the resulting mixing \cite{caulfield_layering_2021,couchman_mixing_2023, riley_effect_2023}. Our approach here will thus be to classify turbulence in terms of the nature of observed density interfaces using a robust processing pipeline starting from raw experimental measurements.

In Sec.~\ref{sec:the-dataset} we review the experimental dataset and its previous human classification. In  Sec.~\ref{sec:detection-morph-density-int}, we describe our new data-driven methodology for automatically discovering distinct turbulent clusters in a low-dimensional phase space. 
Our results are provided in Sec.~\ref{sec:physical_interp}. We first give a physical interpretation of the identified clusters in terms of their characteristic density interface morphology and coherent structures, and highlight how they compare to the human-identified regimes. We then study the transitions of cluster prevalence in the space of input parameters, as well as different types of intermittent behaviors. Finally, we conclude in Sec.~\ref{sec:ccl} and suggest avenues of future exploration.

\section{Experimental dataset and human classification} \label{sec:the-dataset}

\subsection{The Stratified Inclined Duct (SID) setup}

\begin{figure}
\begin{minipage}{0.47\textwidth}
\flushleft (a) The SID setup 
\end{minipage}
\qquad
\begin{minipage}{0.47\textwidth}
\flushleft (b) Dataset and human classification into regimes \\
\end{minipage}
\\
\vspace{0.2cm}
\begin{minipage}{0.47\textwidth}
\includegraphics[width=0.98\linewidth]{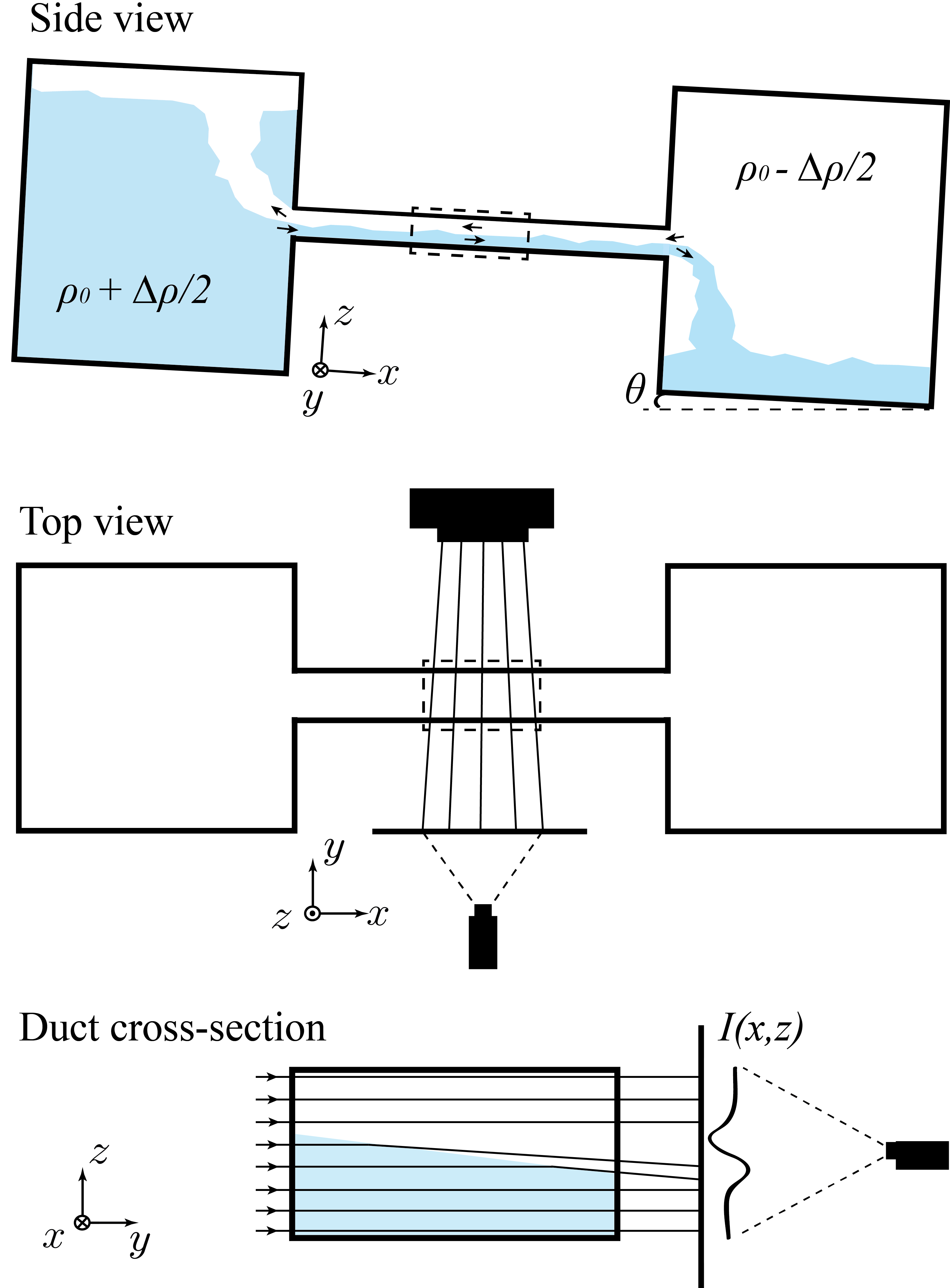}
\end{minipage}
\qquad
\begin{minipage}{0.47\textwidth}
\includegraphics[width=0.98\linewidth]{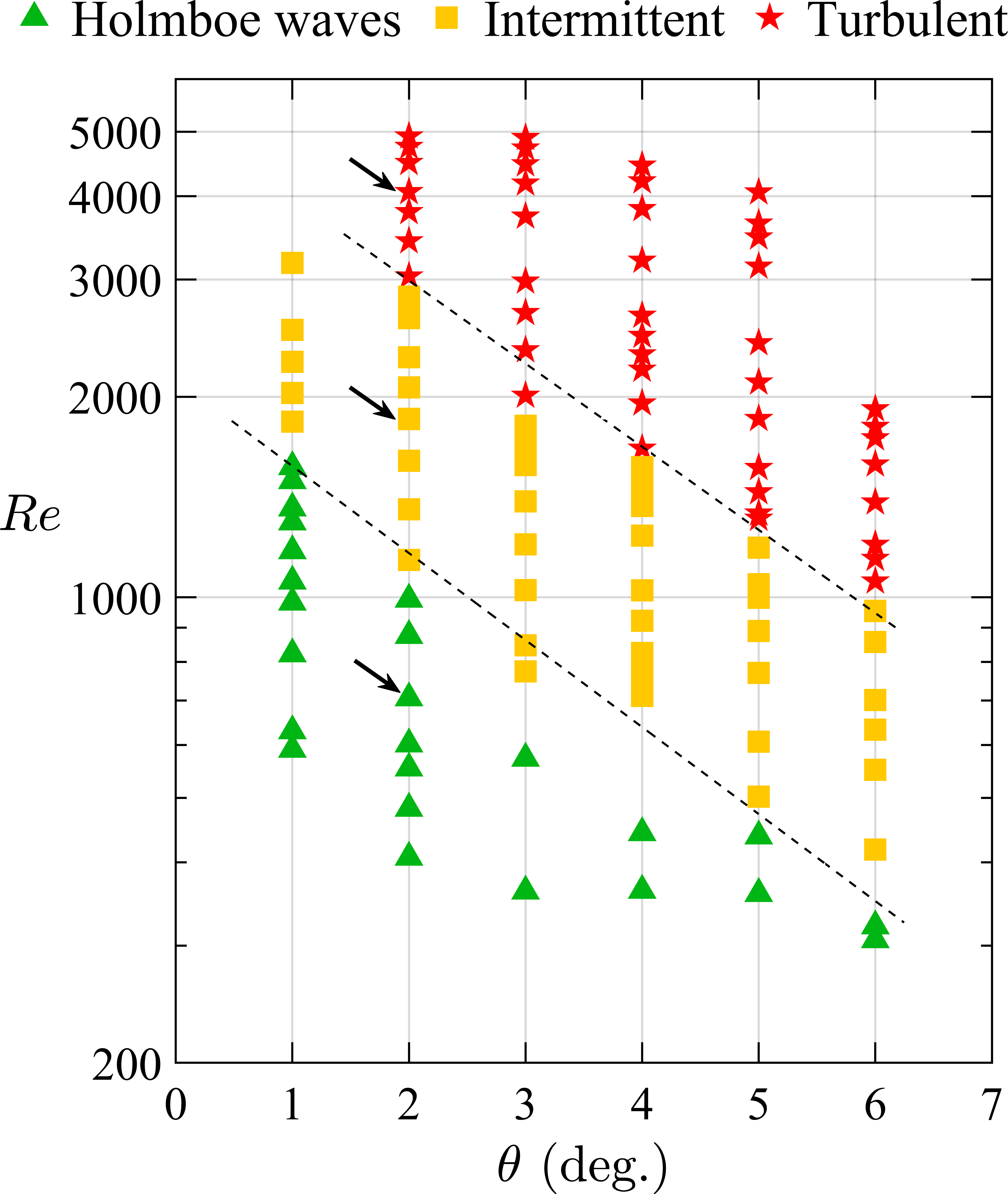}
\end{minipage}
\vspace{0.5cm}
 \flushleft (c) Sample shadowgraph frames \\
\centering
\includegraphics[width=0.7\linewidth]{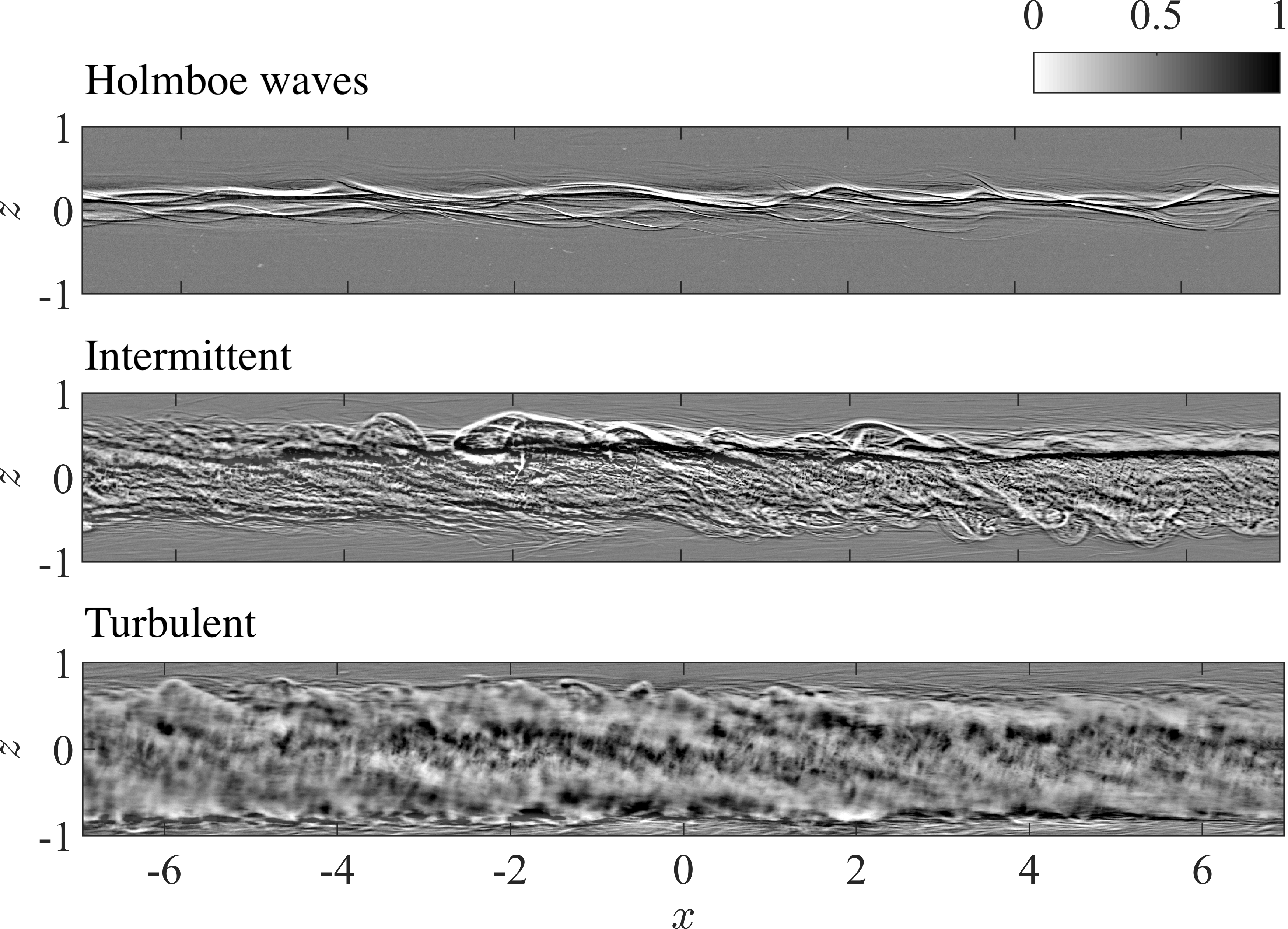}
\caption{Experimental data. (a) Sketch of the setup and shadowgraph measurements used to visualize the density field. (b) The 113 experiments in the space of input parameters $(Re,\theta)$ labelled with their human classified regime. The dashed lines roughly highlight regime transitions. Three arrows indicate the locations of the sample shadowgraph frames $\tilde{I}(x,z)$ presented in (c). The original images are  $\approx$ 3400 $\times$ 450 pixels.}
\label{fig:setup}
\end{figure}

\textit{Principle and geometry} --  The SID setup is sketched in Fig.~\ref{fig:setup}a, consisting of a long rectangular duct connecting two large, closed reservoirs filled with salt (sodium chloride) solutions of different densities $\rho_0 \pm \Delta \rho/2$. The duct is $l=2000$ mm long, $h=50$ mm tall (streamwise aspect ratio $l/h=40$), $w=100$ mm wide (spanwise aspect ratio $w/h=2$), and each reservoir has a volume $V=400$ liters. As the gates isolating the duct from the reservoirs are opened, a two-layer exchange flow through the duct develops. The hydrostratic pressure differential in the reservoirs caused by the reduced gravity $g'=\Delta\rho/\rho_0$ results in a (baroclinic) pressure gradient of opposite sign on either side of the neutral $\rho=\rho_0$ interface, driving the flow with a layer-averaged velocity $\pm u/4 = \sqrt{g'h}/2$ where $u$ is the maximal peak-to-peak velocity scale. The flow can be further energized by inclining the entire apparatus by a small tilt angle $\theta>0$. This accelerates the bottom layer of denser fluid downhill, and the upper layer of buoyant fluid uphill, more than they would under the pressure gradient alone. We align the streamwise axis $x$ along the duct, tilting the $z$ axis by an angle $-\theta$ with respect to the true vertical (opposite to the direction of gravity). The apparatus used here differs from previous two generations  (the first being used in \cite{meyer_stratified_2014} and the second being used in \cite{lefauve_structure_2018,lefauve_regime_2019,partridge_versatile_2019,lefauve_buoyancy_2020,lefauve_experimental1_2022,lefauve_experimental2_2022,jiang_evolution_2022,duran-matute_regime_2023}) in that (i) the duct is located outside rather than inside the reservoirs resulting in cleaner visualizations of the flow; (ii) the reservoirs are larger and closed by rigid lids (there are no free surfaces); (iii) the entire duct-reservoirs dumbbell assembly is tilted at once; and (iv) the duct connects to the reservoirs smoothly with trumpet-shaped ends (adding an extra length of 10~\%).

\textit{Dimensional analysis} -- We non-dimensionalise lengths by $h/2$, velocities by $u/2 = \sqrt{g'h}$ and time by $h/u$, and place the origin of the coordinate system at the center of the duct. The duct volume is thus $(x,y,z)\in [-40,40]\times [-2,2]\times [-1,1]$. The three non-dimensional dynamical parameters are (i) the tilt angle $\theta$ (variable from one experiment to the next); (ii) the Reynolds number $Re=uh/(4\nu) = \sqrt{g'h}h/(2\nu)$ (variable through $\Delta\rho/\rho_0$ in the range $Re=300-5000$); and (iii) the Prandtl number $Pr=\nu/\kappa=700$ (fixed). In our experiments, we take the kinematic viscosity of water as $\nu=1.05\times10^{-6}$ m$^2$ s$^{-1}$ and the molecular diffusivity of salt as $\kappa= 1.5\times 10^{-9}$ m$^2$ s$^{-1}$.  

\textit{Tilt and flow regimes} --  Previous SID experiments, simulations, and two-layer shallow water wave theory showed that the mean exchange flow rate through the duct was  bounded by the  non-dimensional layer-averaged speed $0.5$ as a consequence of `hydraulic control' \citep{meyer_stratified_2014,atoufi_stratified_2023}. This means that the additional power input caused by the tilt $\theta$ cannot be balanced by the viscous dissipation of a faster laminar flow, and must instead be balanced by increasingly dissipative flow structures and interfacial mixing. This causes increasingly turbulent flow regimes with increased $\theta$ (energizing the flow) and with increased $Re$ (reducing the weight of viscous dissipation), as demonstrated in Fig.~\ref{fig:setup}b. We return to a more detailed discussion of these regimes in Sec. \ref{sec:human-class}.  

\textit{Long time series} --  The out-of-equilibrium sheared stratified turbulence in SID persists until each reservoir has been filled with outflowing fluid to mid-level, which takes approximately $V/(h^2w)=1600$ advective time units (A.T.U.), i.e. a fixed non-dimensional time set by the setup geometry. What makes SID particularly valuable is its ability to collect such long time series  at high $Re=O(10^3)$ and $Pr=O(10^3)$, a region of parameter space relevant to salinity- or turbidity-driven environmental and geophysical flows, but currently prohibitive to direct computation. 

\textit{Broader significance} --  From a dynamical systems point of view, SID is also valuable in that the stabilizing effects of density stratification give rise to a richer set of coherent structures, intermittency and transitions at higher $Re$ than in other canonical, unstratified flows (e.g., \cite{deusebio_intermittency_2015}). The discovery of new, high-$Re$ building blocks for the skeleton of stratified turbulence is highly relevant to the broader modeling of multi-physics (e.g. rotating, multiphase, or magnetohydrodynamic) turbulence.

\subsection{Shadowgraphs of the density field}

\textit{Principle} --  The density field is visualized using a shadowgraph technique that involves shining light through the duct and measuring its refraction (Fig.~\ref{fig:setup}a). Approximately parallel light rays produced by a slide projector travel through the duct along the spanwise $y$-direction and are projected onto a semi-transparent screen. Any variations in the curvature (normal to the rays) of the local perturbation density field $\rho(x,y,z,t)-\rho_0$, and thus of the refractive index field $N(x,y,z,t)$, causes the rays to focus or defocus, varying the light intensity that reaches the screen. In the limit of weak variations, the intensity of the image formed and recorded by video camera is (see e.g. \cite[\S~2.1]{lefauve_waves_2018})
\begin{equation}\label{eq:sg}
    I(x,z,t) = \beta \, I_0(x,z) \int_{-2}^{2} \Big(\frac{\partial^2}{\partial x^2}+\frac{\partial^2}{\partial z^2}\Big)\rho(x,y,z,t) \ dy.
\end{equation}
Here $\beta$ depends on $(\rho_0/N_0)\partial N/\partial\rho$ and the experimental geometry, and $I_0$ is the (approximately) uniform background intensity of the illumination. The shadowgraph signal $I$ is thus particularly well suited to detecting density interfaces, the structures of interest here for distinguishing between different turbulent regimes. A sharp density gradient $\partial \rho/\partial z$ will result in an intensity $I$ having a low (dark) and high (bright) peak on either side of it.

\textit{Acquisition} -- We carried out 113 individual experiments at six different tilt angles $\theta$ equally spaced between $1^\circ$ and $6^\circ$. Each campaign was run at a fixed $\theta$ by initially filling the left reservoir with brine and the right reservoir with fresh water, resulting in a large $\Delta\rho/\rho_0$ and thus a large $Re\approx 2000-5000$. The duct was opened to start the exchange flow, and time was counted ($t=0$) from the moment the gravity currents originating from either ends of the duct reached the centre of the duct ($x=0$). Shadowgraph movies were then recorded with a video camera tilted at the same angle $\theta$ as the setup to record natively in the $(x,z$) coordinate system, covering the full internal height of the duct (50 mm) and a width $\approx 325-425$ mm (depending on the campaign), centred at $x=0$.  The frame rate was set between 5 and 100 fps depending on the speed of the flow, to achieve a typical frame spacing of 0.1 non-dimensional A.T.U., and 400-600 A.T.U. were typically recorded (recalling that time is non-dimensionalized in each experiment by $h/u=h/(2\sqrt{g'h})$).  The experiment was then stopped by closing off the duct at both ends and  mixing the fluid in both reservoirs, thus reducing $\Delta\rho/\rho_0$ and $Re$. The next experiment at a lower $Re$ was then started, recorded, and so forth, until the lowest $Re\approx 300-600$. This allowed us to cover the $(Re,\theta)$ space of Fig.~\ref{fig:setup}b, consisting of 15 experiments at $\theta=1^\circ$; 22 at $\theta=2^\circ$; 19 at $\theta=3^\circ$; 21  at $\theta=4^\circ$; 20  at $\theta=5^\circ$; and 16 at $\theta=6^\circ$. The light intensity across all videos was normalized to yield $\tilde{I}(x,z,t)$, the initial transient ($t<100$) discarded, and the temporal resolution coarsened, as further described in Appendix~\ref{sec:appendix-data-processing}. Three example frames are shown in Fig.~\ref{fig:setup}c. The final dataset consists of 113 movies totalling 50155 frames, giving an average of 444 frames per experiment with a typical frame-to-frame spacing of 1 A.T.U.

\subsection{Human classification into flow regimes} \label{sec:human-class}

Qualitative flow visualizations, including shadowgraph movies, have previously been used to classify SID measurements into four flow regimes: laminar (L), Holmboe waves (H), intermittent (I), and turbulent (T), as we have done for our current dataset in Fig.~\ref{fig:setup}b. Such a manual classification was first introduced by Macagno \& Rouse \cite{macagno_interfacial_1961} (hereafter MR61), and subsequently rediscovered (without knowledge of MR61) by Meyer \& Linden \cite{meyer_stratified_2014} (hereafter ML14), using almost identical descriptions. We quote and compare their descriptions of the four qualitatively different flow regimes below:
\begin{itemize}
    \item[(L)]  ``uniform laminar motion with straight streamlines'' (MR61) and ``an undisturbed density interface separating the two layers'' (ML14); 
    \item[(H)]  ``laminar motion with regular waves'' (MR61) and ``the flow is wave-dominated and exhibits Holmboe modes on the interface, with characteristic cusp-like wave breaking'' (ML14);
    \item[(I)] ``incipient turbulence, with waves which break and start to show irregularity and randomness'' (MR61) and  ``intermittent state, which exhibits a rich range of spatio-temporal behavior and an interfacial region that contains features of Kelvin–Helmholtz-like structures and of the other two lower-dissipation states: thin interfaces and Holmboe-like structures''(ML14);
    \item[(T)] 
``pronounced turbulence and active mixing across the interface'' (MR61) and 
 ``turbulent high-dissipation interfacial region typically containing Kelvin–Helmholtz-like structures sheared in the direction of the mean shear and connecting both layers''  (ML14).
\end{itemize}

Fig.~\ref{fig:setup}c shows examples of a shadowgraph frame in the H, I and T regimes. Note that the dataset in this paper does not contain flows in the L regime (found at lower values of $Re,\theta$ than considered in Fig.~\ref{fig:setup}b) because their flat, sharp interface and steadiness render them uninteresting to our analysis. The main difference between the I and T regimes is that the latter never relaminarises.


\textit{Limitations} -- As explained in the Introduction, the classification of an entire movie into a single regime causes three problems: (i) arbitrariness and inconsistency; (ii) implicit assumption of sharp transitions; (iii) neglect of spatial and temporal complexity. For example, the temporal variations between various quasi-laminar wave structures and more turbulent structures is essential to the fascinating intermittent regime, which sometimes exhibits quasi-periodic laminar-turbulent cycles with a wide range of flow structures, and for which a probabilistic description appears needed. Three-dimensional velocity and density experimental data have also revealed different `flavours' of turbulence, even at comparable values of turbulent kinetic energy dissipation proportional to $Re \, \theta$ \cite{lefauve_regime_2019,lefauve_experimental2_2022}, with low-$\theta$ flows having more extreme enstrophy but less overturning events than high-$\theta$ flows \cite{lefauve_experimental1_2022}. Such valuable insight cannot easily be drawn by eye from numerous shadowgraph movies. This motivates the need for an automated classification based on physically-interpretable coherent structures, which we formalise and apply next.

\section{Dimensionality reduction and classification} \label{sec:detection-morph-density-int}

\subsection{Overview of the method}

\begin{figure}
\centering
\includegraphics[width=0.93\linewidth]{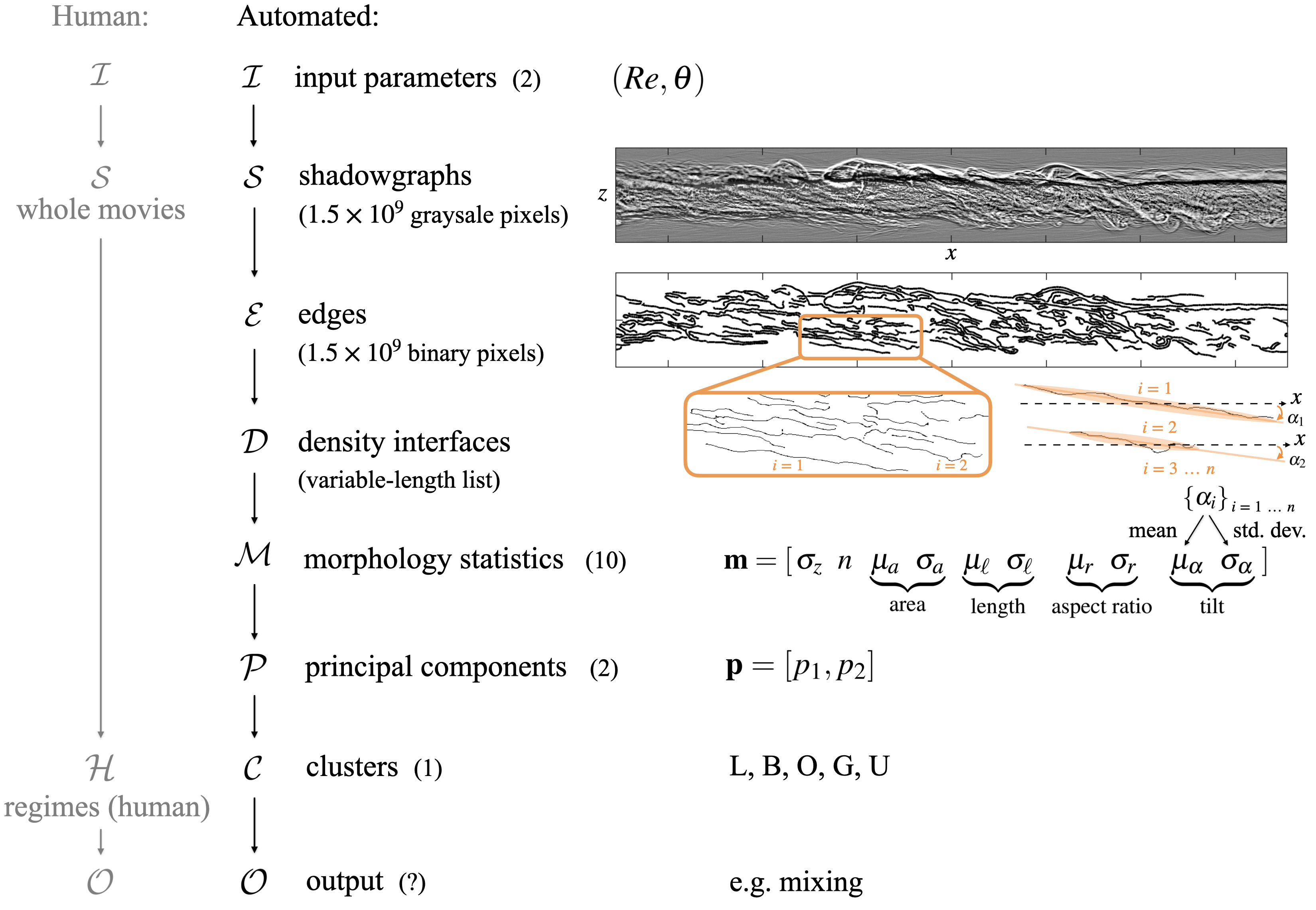}
\caption{Dimensionality reduction pipeline used to assign a cluster label to each shadowgraph frame. The spaces and successive mappings of the new automated classification approach (second column from left, in black) are compared to the traditional human approach (left, in gray). The bracketed numbers indicate the dimension of each space, highlighting a dramatic reduction in dimensionality.}
\label{fig:pipeline}
\end{figure}

This section introduces the automated pipeline that takes an entire dataset of shadowgraph images, collected across the two-dimensional parameter space spanned by Reynolds number $Re$ and tilt angle $\theta$, and determines a natural grouping of these data into distinct clusters in another low-dimensional space, where the number of clusters and their properties are initially unknown. Fig.~\ref{fig:pipeline} summarises our approach, and each step is detailed in the following sections. 

In summary, each shadowgraph image is first transformed into a collection of binary edges delinating the locations of sharp density gradients (see \S~\ref{sec:edge}). A variety of geometrical properties of each interface are then computed and their statistics  are used to form a low-dimensional vector of characteristics (\S~\ref{sec:CC}). A principal component analysis is then performed on the entire set of morphology vectors, demonstrating that the dominant trends in the morphology statistics data may be captured in a two-dimensional subspace (\S~\ref{sec:PCA}). The two-dimensional vectors representing all shadowgraph frames are then automatically classified (\S~\ref{sec:Clust}), revealing five clusters with distinct properties. We interpret these clusters in \S~\ref{sec:physical_interp} by analyzing the inverse mappings ($\mathcal{C} \rightarrow \mathcal{D},\mathcal{E},\mathcal{S},\mathcal{I}$), comparing the clusters to human-identified regimes ($\mathcal{C} \rightarrow \mathcal{H}$), and studying the temporal trajectories within or between clusters.


We note that the traditional human classification approach (sketched in gray, left-column Fig.~\ref{fig:pipeline}) mapped a collection of frames (a movie) directly to the one-dimensional space of regimes $\mathcal{H}$, containing three possible values: Holmboe wave (H), intermittent turbulence (I), and sustained turbulence (T). By contrast, our automated classification approach provides a series of objective, repeatable mappings which remain easily interpretable. This interpretability distinguishes our approach from other data-driven approaches relying on relatively opaque algorithms such as deep neural networks (e.g. autoencoders).

\subsection{Edge detection ($\mathcal{S}\rightarrow \mathcal{E}$)} \label{sec:edge}

\textit{Canny edge detection} -- A Canny edge-detection algorithm \cite{canny_computational_1986}, implemented using Matlab's function \texttt{edge}, was applied to each shadowgraph frame  (containing $\approx 1.5$ million pixels) in order to transform the grayscale shadowgraph image of the density field to a binary image delineating the positions of sharp density gradients. 
The Canny algorithm works by first lightly smoothing the grayscale image (normalized to have values between 0 and 1)  with a Gaussian filter (here with an isotropic standard deviation of $5$ pixels) and then numerically computing the gradient of the filtered image. 
In order to pick out edges, two thresholds are considered. If a pixel gradient is higher than the upper threshold (set here at 0.5), the pixel is accepted as an edge. If a pixel gradient is below the lower threshold (set here at 0.05), then it is rejected.  If a pixel gradient is between the two thresholds, it is accepted only if it is connected to a pixel that is above the upper threshold (an edge). This double thresholding improves the detection of true weak edges (avoiding true negatives) and robustness to noise (avoiding false positives). Identical thresholds were used for all frames, and the detected contours were only weakly dependent on the thresholds used.

\begin{figure}
\centering
\includegraphics[width=0.7\linewidth]{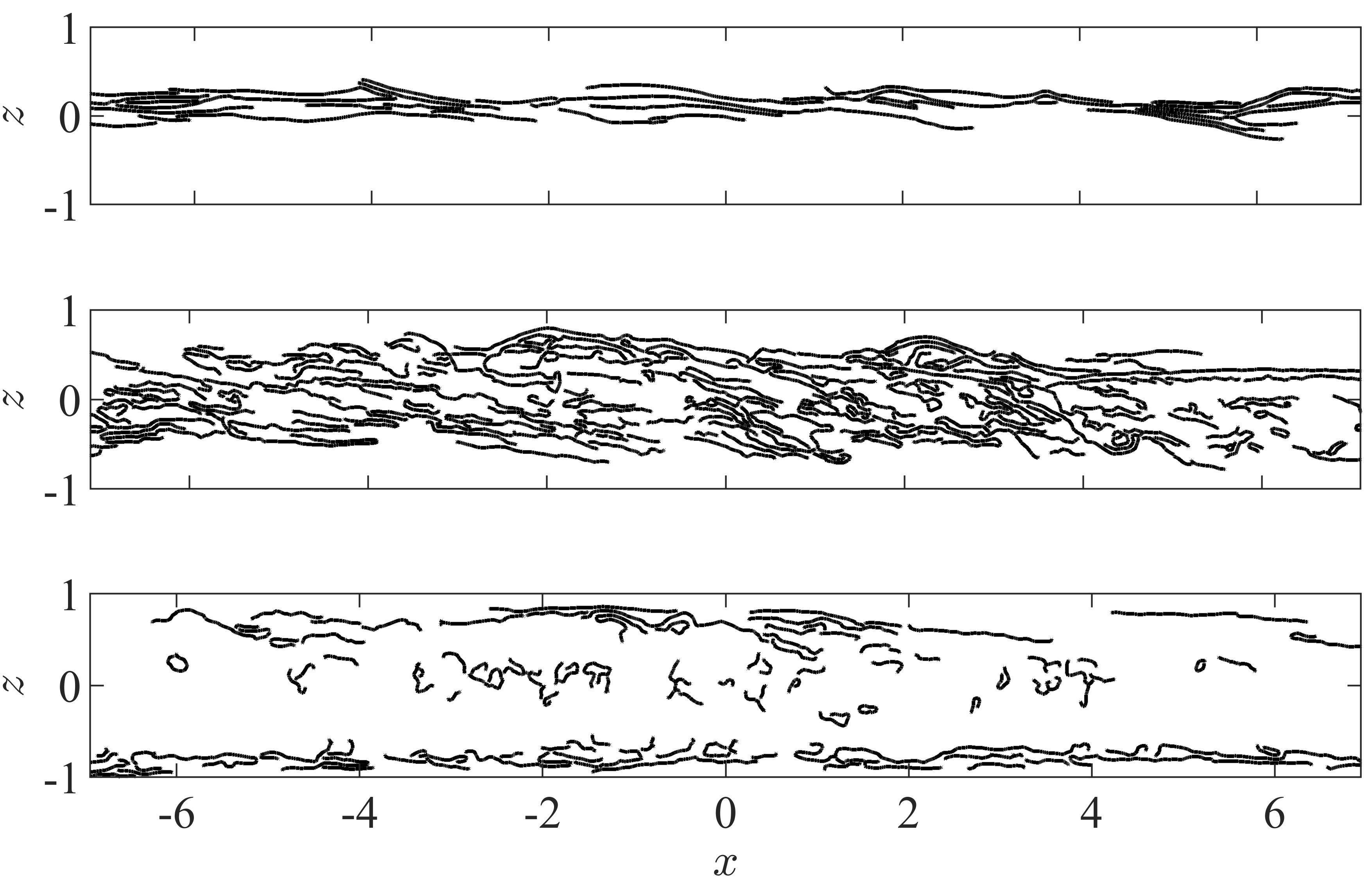}
\caption{Sample binary edge frames ($\mathcal{E}$ space), delineating the locations of sharp density interfaces (black), mapped from the shadowgraphs ($\mathcal{S}$ space) of figure~\ref{fig:setup}c in the three human-classified regimes: Holmboe (H, top), intermittent (I, middle), and turbulent (T, bottom). Edges are here rendered much thicker than detected for better visualization. }
\label{fig:edges}
\end{figure}

\textit{Results} -- Fig.~\ref{fig:edges} shows examples of the binary edge images $e(x,z) \in \mathcal{E}$  corresponding to the shadowgraphs $\tilde{I}(x,z) \in \mathcal{S}$ of Fig.~\ref{fig:setup}c. Typically pixel-thin, the edges are rendered here with much thicker black contours for better visualization. The Holmboe (H) flow (top) exhibits a relatively sharp but undulating density interface with cusped waves, and typically multiple edges, or filaments, stacked on top of one another. The thicker layer of intermediate density of the intermittent (I) flow (middle) results in more numerous edges, some of which are shorter along $x$ and more tilted with respect to the $x$ axis. The even thicker intermediate layer in the turbulent (T) flow is so turbulent that few edges are detected within it (at low $|z|$), as the three-dimensional nature of the turbulence blurs the resulting shadowgraph image, but some edges are detected at the upper and lower edges of the frame ($\left|z\right|\approx1$). The edges within the intermediate layer tend to be relatively short and tilted away from the horizontal, revealing instability and overturning motions, whereas the edges on either side of it are longer and flatter, revealing more quiescent stable interfaces. This is consistent with gradient Richardson number profiles $Ri_g(z)$ \cite[Fig. 4]{lefauve_regime_2019} (quantifying the competition between destabilizing shear and stabilizing stratification); low $Ri_g\approx 0.15$ were found throughout the turbulent layer while higher $Ri_g>1$ were found at the interfaces on either side.

\subsection{Discrete density interfaces and morphology statistics ($\mathcal{E}\rightarrow \mathcal{D}\rightarrow\mathcal{M}$)} \label{sec:CC}

\textit{Connected components} -- Having generated a binary image of sharp density interfaces, we now wish to quantify the morphological properties of each connected edge. This is achieved by first applying a connected component algorithm to the binary image, using Matlab's function \texttt{bwconncomp}. A connected component is defined here as a set of pixels that are connected on any of their four sides or four corners (often referred to as a two-dimensional pixel connectivity of eight). In other words, two adjoining pixels are considered part of the same density interface if they are connected along the horizontal, vertical, or diagonal direction. This algorithm returns $n$ discrete density interfaces for a given frame, characterized by a list of pixels belonging to each.

\textit{Edge properties} -- We then compute the morphology of each density interface within a given frame using Matlab's function \texttt{regionprops} with the following  four arguments: `Area', returning the actual number of pixels in the interface; `MajorAxisLength' and `MinorAxisLength', returning the length (in pixels) of the major and minor axes, respectively, of the ellipse that has the same normalized second central moments as the interface; and `Orientation', returning the angle between the $x$-axis and the major axis of this ellipse (with positive angles denoting anticlockwise rotations). The areas and lengths are then converted from pixels to non-dimensional units (like the axes of Fig.~\ref{fig:edges}) so that all frames can be compared consistently with meaningful physical values. This yields, for each frame, a list of $n$ non-dimensional density interface areas $\{a_i\}_{i=1,\ldots,n}$, lengths $\{\ell_i\}$ (from the major axis), aspect ratio $\{r_i\}$ (ratio of major to minor axes), and tilts $\{\alpha_i\}$ ($0$ is aligned with $\hat{x}$, and $\pm 90^\circ$ are aligned with $\pm \hat{z}$ respectively). These lists can be seen as belonging to the space $\mathcal{D}$, describing the morphological properties of the edges within each frame, which can be conveniently visualized by plotting histograms of $a,\ell,r,\alpha$ (which we will show in \S~\ref{sec:results-C-M}).


\textit{Morphology vector} -- Finally, in order to distil the properties of the multiple edges within a frame into a single vector, we compute the mean ($\mu$) and standard deviation ($\sigma$) of the distributions $\{a_i\},\{ \ell_i \},\{r_i \}, \{\alpha_i\}$, characterizing the centre of mass and moment of inertia of their histograms. To this eight-dimensional vector, we added two further components: the number of interfaces $n$, and the moment of inertia of the edges with respect to the vertical coordinate $\sigma_z=\sqrt{\langle \langle e\rangle_x \, (z-\langle  e\rangle_{x,z} )^2 \rangle_z}$, where larger values indicate a greater spread of density interfaces around the mean position. This yielded the following 10-dimensional vector of morphology statistics for each frame:
\begin{equation} \label{eq:x}
    \mathbf{m} = [\, \sigma_z  \  \ n \ \ \underbrace{\mu_a  \ \ \sigma_a}_{\text{area}} \ \ \underbrace{\mu_\ell \ \ \sigma_\ell}_{\text{length}} \  \ \underbrace{\mu_r  \  \ \sigma_r}_{\text{aspect ratio}} \  \ \underbrace{\mu_\alpha \ \ \sigma_\alpha}_{\text{tilt}} \ ] \ \in \ \mathcal{M}.
\end{equation}
The 50155 row vectors generated from all frames are then arranged into a single tall, skinny matrix denoted $\mathbf{M} \in \mathbb{R}^{50155\times10}$. 


\subsection{Principal Component Analysis ($\mathcal{M}\rightarrow \mathcal{P}$)} \label{sec:PCA}

\textit{Motivation} -- Before clustering, it is worth exploiting potential correlations between the morphological features within $\mathcal{M}$ (see definition \ref{eq:x}) by performing a principal component analysis (PCA) \cite{brunton_data-driven_2019}. Our goal is to further reduce the dimensionality of the data, to assist with the interpretation of the clustering results and to improve the effectiveness of the clustering algorithm due to the `curse of dimensionality', i.e. the dramatic increase in the volume of the clustering space with  increasing dimensions. 

\begin{figure}
\vspace{-0.2cm}
\flushleft(a) Correlations between morphology statistics  \\ \vspace{0.2cm}
\centering
\includegraphics[width=0.78\linewidth]{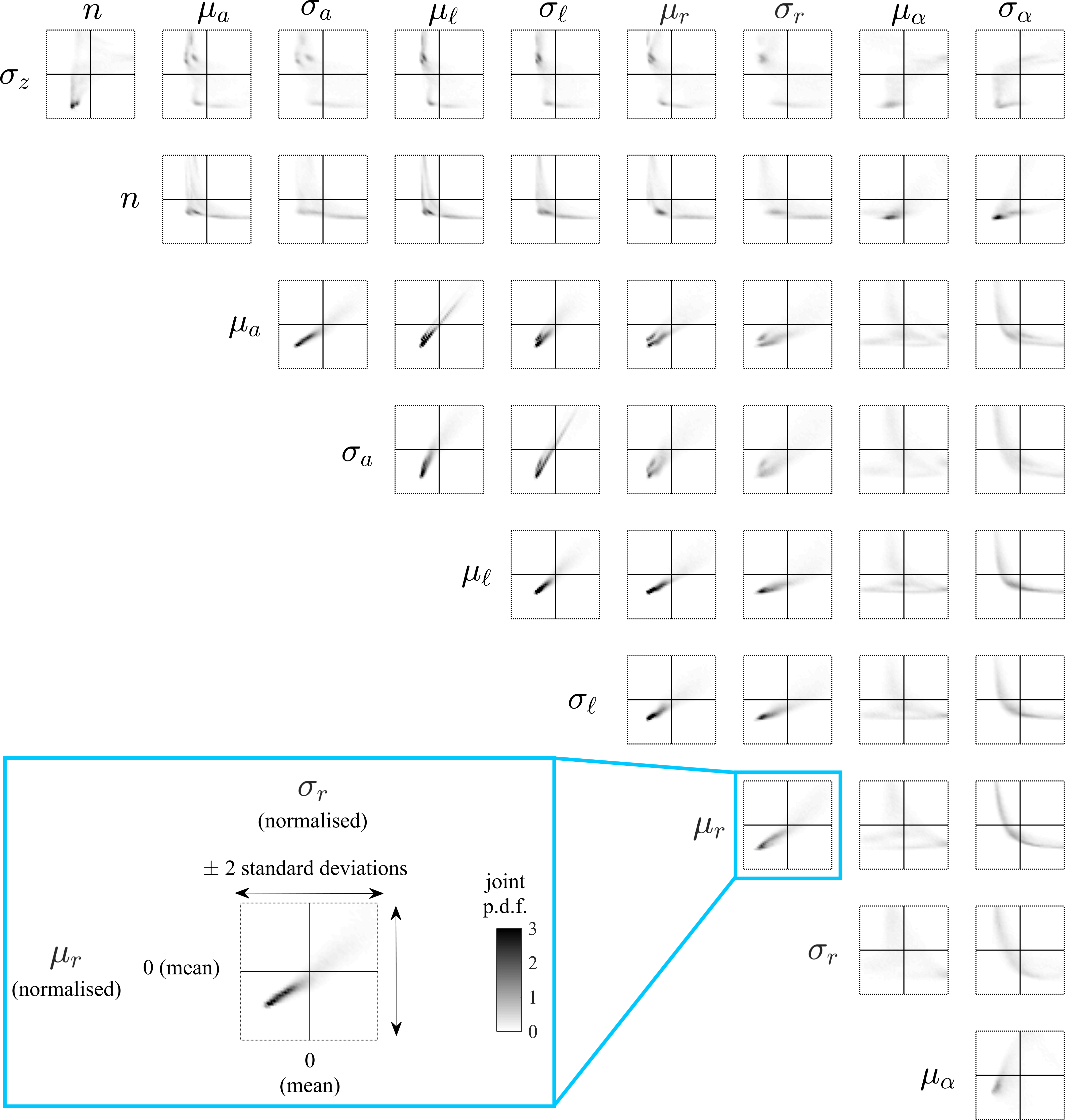}
\begin{minipage}{0.43\textwidth}
\flushleft (b) Variance and $k=2$ truncation
\end{minipage}
\begin{minipage}{0.54\textwidth}
\flushleft (c) The $\mathcal{M}\rightarrow \mathcal{P}$ mapping \\
\end{minipage}
\begin{minipage}{0.43\textwidth}
\vspace{0.4cm}
\includegraphics[width=\linewidth]{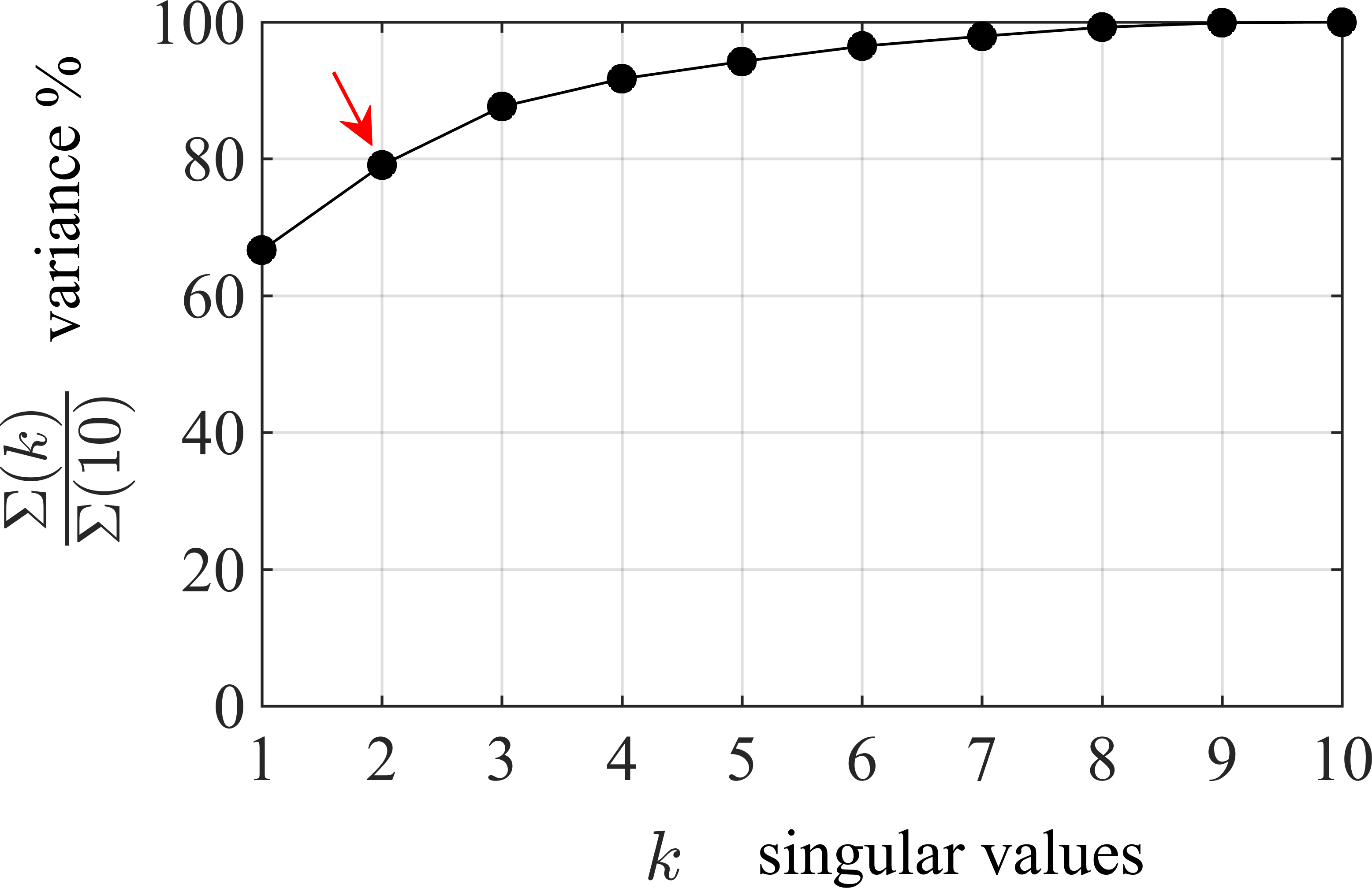}
\end{minipage}
\hspace{0.1cm}
\begin{minipage}{0.54\textwidth}
\includegraphics[width=\linewidth]{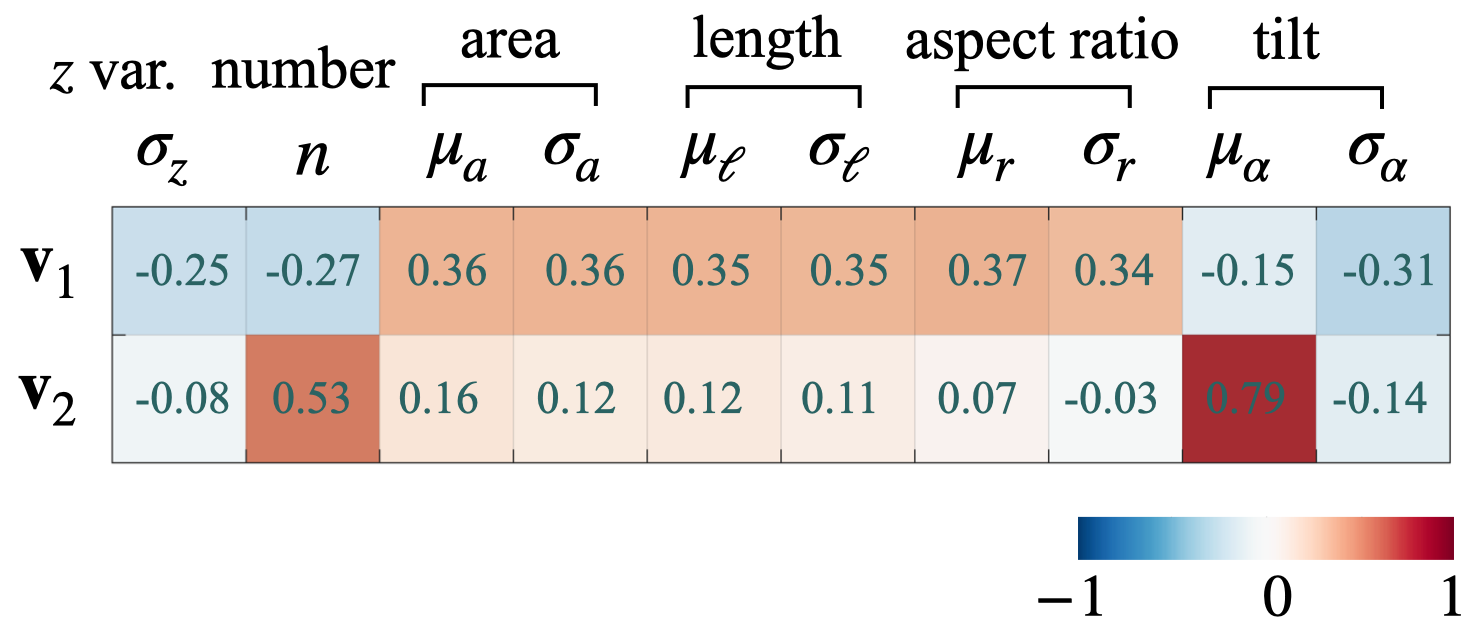}
\end{minipage}
\vspace{0cm}
\caption{(a) Joint p.d.f.s of the 10 statistics characterizing the morphology of density interfaces (see Eq. \ref{eq:x}). The example inset (in blue) shows correlation between $\mu_r$ (the mean aspect ratio of density interfaces) and  $\sigma_r$ (the variability in their aspect ratio). (b)  Cumulative variance captured by the principal components (or singular values); the red arrow highlights that $k=2$ captures 79~\% of the total. (c) Coordinates of the first two principal unit vectors $\mathbf{v}$ in terms of the original basis of morphology statistics. }
\label{fig:correlation_PCA}
\end{figure}

\textit{Correlations and rank-two approximation} -- We start by normalizing each column of $\mathbf{M}$ to have zero mean and unit standard deviation, and obtain $\tilde{\mathbf{M}}$. Fig.~\ref{fig:correlation_PCA}a shows the correlation between all $10\times 9/2=45$ pairs of variables, highlighting the structure of the covariance matrix $\widetilde{\mathbf{M}}^T\widetilde{\mathbf{M}}$. We find many direct and inverse correlations, motivating the need to express these data in a basis of linearly uncorrelated variables, called principal components.  We perform the singular value decomposition (SVD) of $\tilde{\mathbf{M}}$ (for more details, see Appendix~\ref{sec:appendix-pca}) and plot in Fig.~\ref{fig:correlation_PCA}b  the cumulative variance  $\Sigma(k) = \sum_{l=1}^k\sigma_l^2$ captured by the first $k$ singular values. We find that the first $k=2$ principal components capture 79~\% of the total variance, and given the convenience of visualizing a two-dimensional phase space, we choose to truncate the SVD at $k=2$ to obtain a new rank-two data matrix $\mathbf{P}\in \mathbb{R}^{50155\times 2}$ containing two-dimensional vectors $\mathbf{p} \in \mathcal{P}$.

\textit{Principal component space} -- Fig.~\ref{fig:correlation_PCA}c plots the coordinates of the two new orthogonal basis vectors $\mathbf{v}_1,\mathbf{v}_2$ spanning $\mathcal{P}$ (pointing in the directions of maximal variance) in the previous basis spanning $\mathcal{M}$. They explicitly show the linear combination of morphology statistics making up each principal component, thereby defining the $\mathcal{M}\rightarrow\mathcal{P}$ map. Simply put, principal vector $\mathbf{v}_{1}$ weighs almost equally the area $a$, length $\ell$, and aspect ratio $s$ with positive values ($\approx 0.4$, in orange) and the remaining statistics with negative values ($\approx -0.3$, in light blue), except the mean tilt $\mu_\alpha$, which is weaker. Principal vector $\mathbf{v}_{2}$, on the other hand, weighs much more heavily this mean tilt and the number of interfaces $n$ (in red).

\subsection{Clustering ($\mathcal{P}\rightarrow \mathcal{C}$)} \label{sec:Clust}

\textit{Algorithm} -- Having reduced the data within each frame of the density field to a point in the two-dimensional space $\mathcal{P}$, we now perform a clustering analysis on the set of all frames. We choose to use the density-based clustering algorithm OPTICS \citep{ankerst_ordering_1999} (`Ordering Points To Identify the Clustering Structure'), recently applied to the discovery of distinct regions of turbulent mixing within an ocean microstructure dataset \cite{couchman_data-driven_2021}.  OPTICS has three main advantages over other algorithms: it determines the optimal number of clusters automatically, it identifies clusters of arbitrary shape and density, and it is robust to noise and outliers \cite{han_cluster_2012}. OPTICS uses a metric called the `reachability distance' $d_{\text{reach}}$ to compute the pairwise distances between the rows of $\mathbf{P}$ (for more details, see Appendix~\ref{sec:appendix-optics}). Small values of $d_{\text{reach}}$ indicate a high local density of frames in $\mathcal{P}$.

\begin{figure}
\centering
\includegraphics[width=0.88\linewidth]{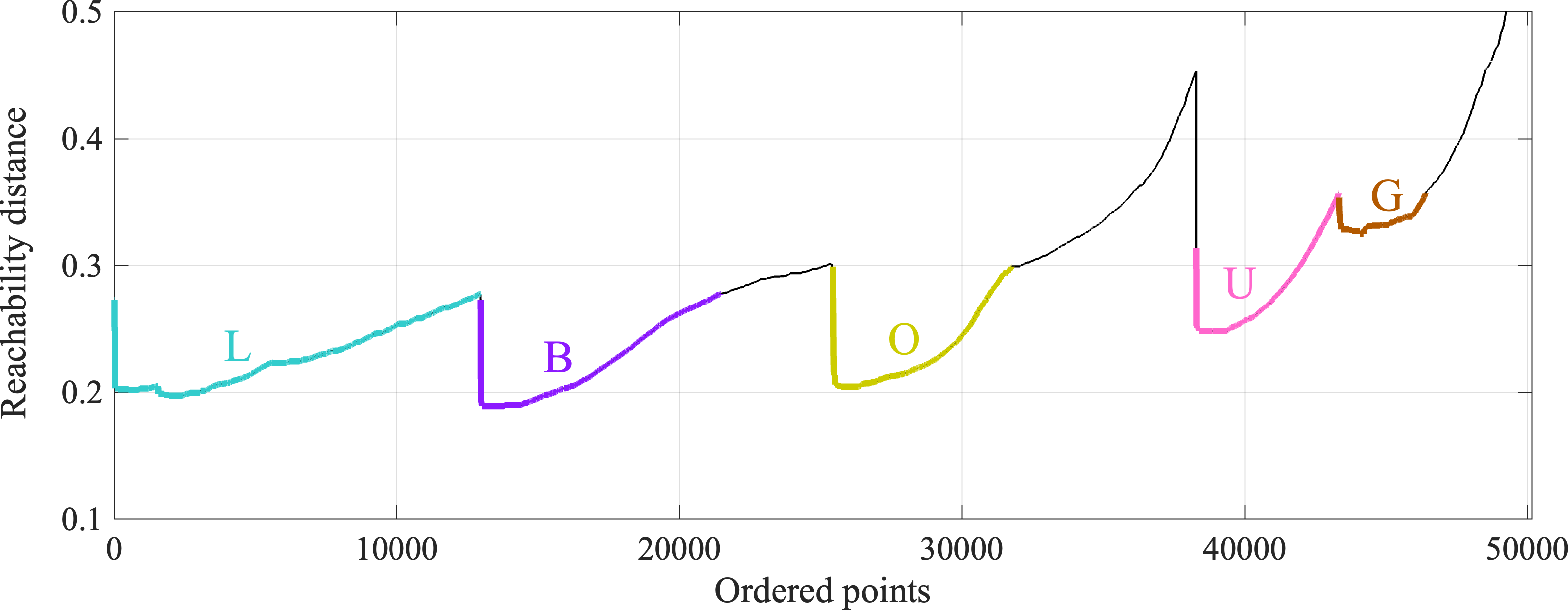}
\caption{The five distinct clusters (colored, labelled L-G) detected by the OPTICS algorithm, corresponding to local minima in the reachability distance (defined in Appendix~\ref{sec:appendix-optics}). Unclustered data are colored black. The  distance monotonically increases above the $y$-axis limit of 0.5 for the last $\approx 1000$ unclustered points.}
\label{fig:reach}
\end{figure}

\textit{Results} -- The output reachability plot is shown in Fig.~\ref{fig:reach}. It is generated sequentially from left to right: starting from an arbitrary row in $\mathbf{P}$, at each step OPTICS moves to the next closest point based on the reachability distance, and plots $d_{\text{reach}}$ at that step.  OPTICS thus steps toward a region with a high local density, as reflected by a steady decrease in $d_{\text{reach}}$.  Having visited each point in this dense region, it then automatically moves to the next closest points  in sparser intervening regions, reflected by a larger $d_{\text{reach}}$, before entering a new locally dense region (if one exists), etc. 

The five valleys in Fig.~\ref{fig:reach} reveal five clusters, with lower minima indicating denser clusters. Using this reachability plot, we manually split and label the five clusters L, B, O, G, U, excluding some local maxima between clusters which we consider unclustered. Overall, 72\,\% of all points belong to clusters (26\,\% in L, 17\,\% in B, 13\,\%  in O, 6\,\% in G, 10\,\% in U) and 28\,\% are unclustered. 
The physical interpretation of these clusters is given in the next section.

\begin{figure}
\centering
\includegraphics[width=0.6\linewidth]{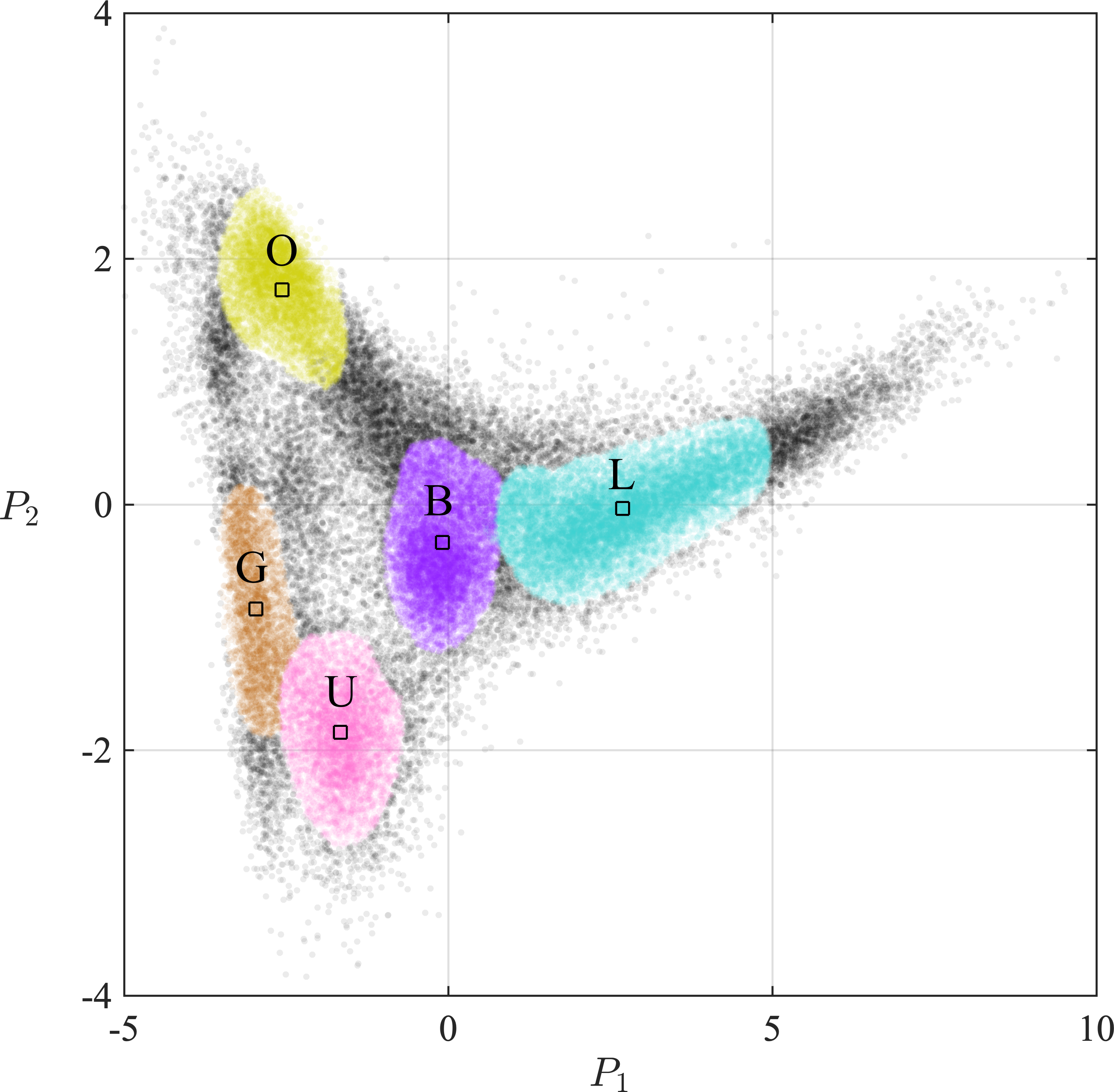}
\caption{Clusters (as identified and colored in Fig.~\ref{fig:reach}) plotted in the space of the two dominant principal components of the density interface morphology statistics.  The relation between the two principal component coordinates and the 10 morphology statistics is illustrated in Fig.~\ref{fig:correlation_PCA}c. }
\label{fig:clusters_PCA}
\end{figure}

\section{Results and physical interpretation} \label{sec:physical_interp}

Having identified five clusters, we now interpret their properties by first analysing their distribution in the clustering space, and then working back to the original shadowgraphs. We then use this insight to build a picture of the dominant dynamics across the parameter space of SID.

\subsection{In terms of principal components ($\mathcal{C}\rightarrow \mathcal{P}$)} 

Fig.~\ref{fig:clusters_PCA} shows the location of the five clusters L, B, O, U, G in the subspace of the first two principal components. Each translucent circle represents one of the 50155 shadowgraph frames of the density field as is colored according to its assigned cluster based on the reachability plot in Fig.~\ref{fig:reach}. Darker shading indicates a greater local density of points. Black circles denote the 28 \% unclustered frames which did not form regions of sufficient local density to be considered part of a cluster.

The data organise into a rough (non-convex) triangle. As the data were originally normalized using a $z$-score (see $\S$ \ref{sec:PCA}), the origin corresponds to the mean of the data, and $P_1$ and $P_2$ measure the number of standard deviations away from the mean in the directions of the respective PCA vectors. 
The vast majority of points belong to the rectangle $[-5,10]\times [-4,4]$, and all five clusters belong to the smaller rectangle $[-4,5]\times[-3,3]$. The five black square symbols denote the centroid of each cluster, which we interpret next by projecting back to the morphology space $\mathcal{M}$.

\subsection{In terms of density interfaces morphology ($\mathcal{C}\rightarrow \mathcal{M}$)} \label{sec:results-C-M}



\begin{figure}
\vspace{-0.7cm}
\flushleft(a) Typical interface morphology of each clusters \\ \vspace{0.2cm}
\centering
\includegraphics[width=0.63\linewidth]{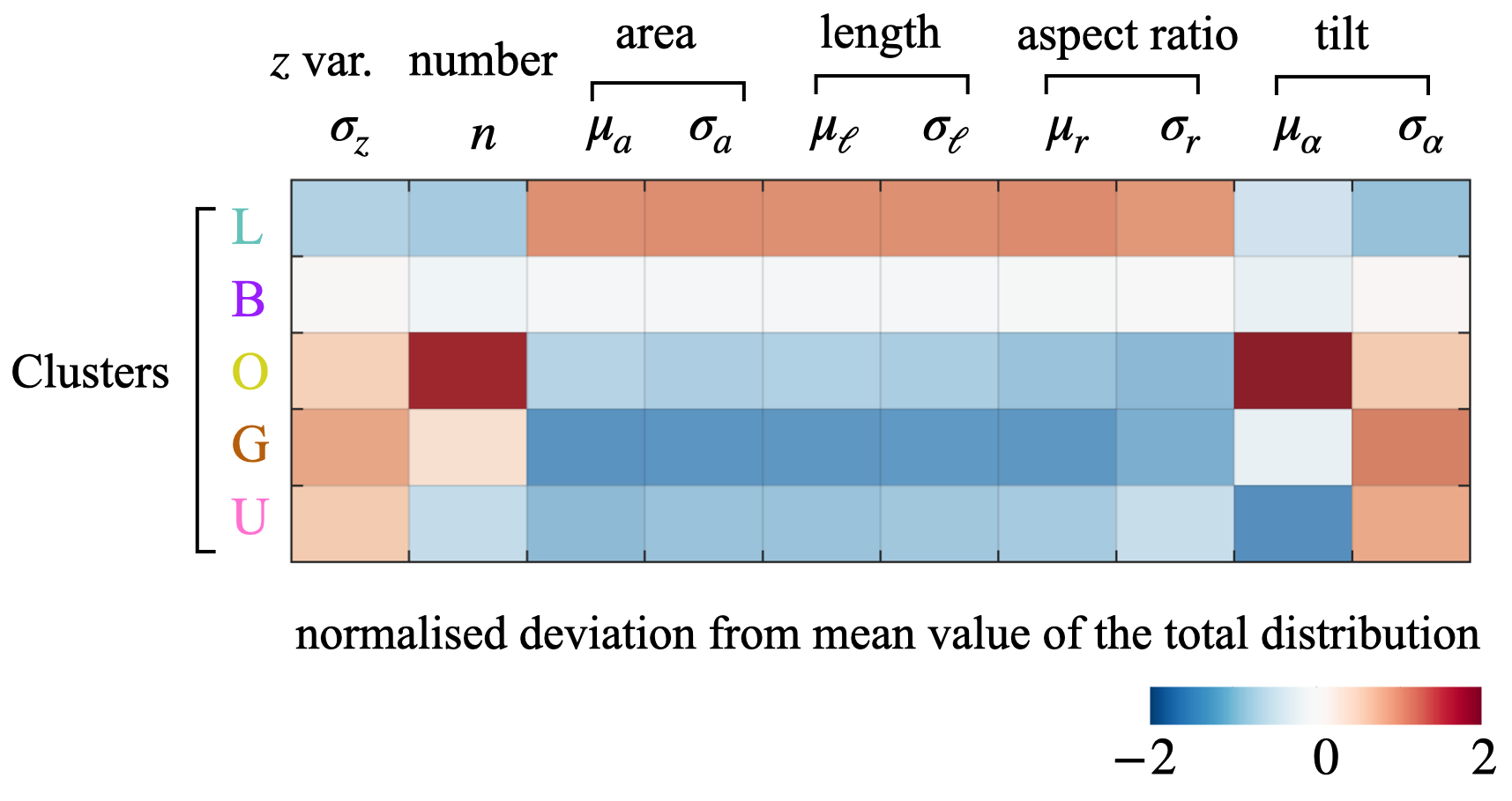}  \vspace{-0.5cm}
\flushleft(b) Distribution of morphology statistics \\ \vspace{0.4cm}
\centering
\includegraphics[width=0.95\linewidth]{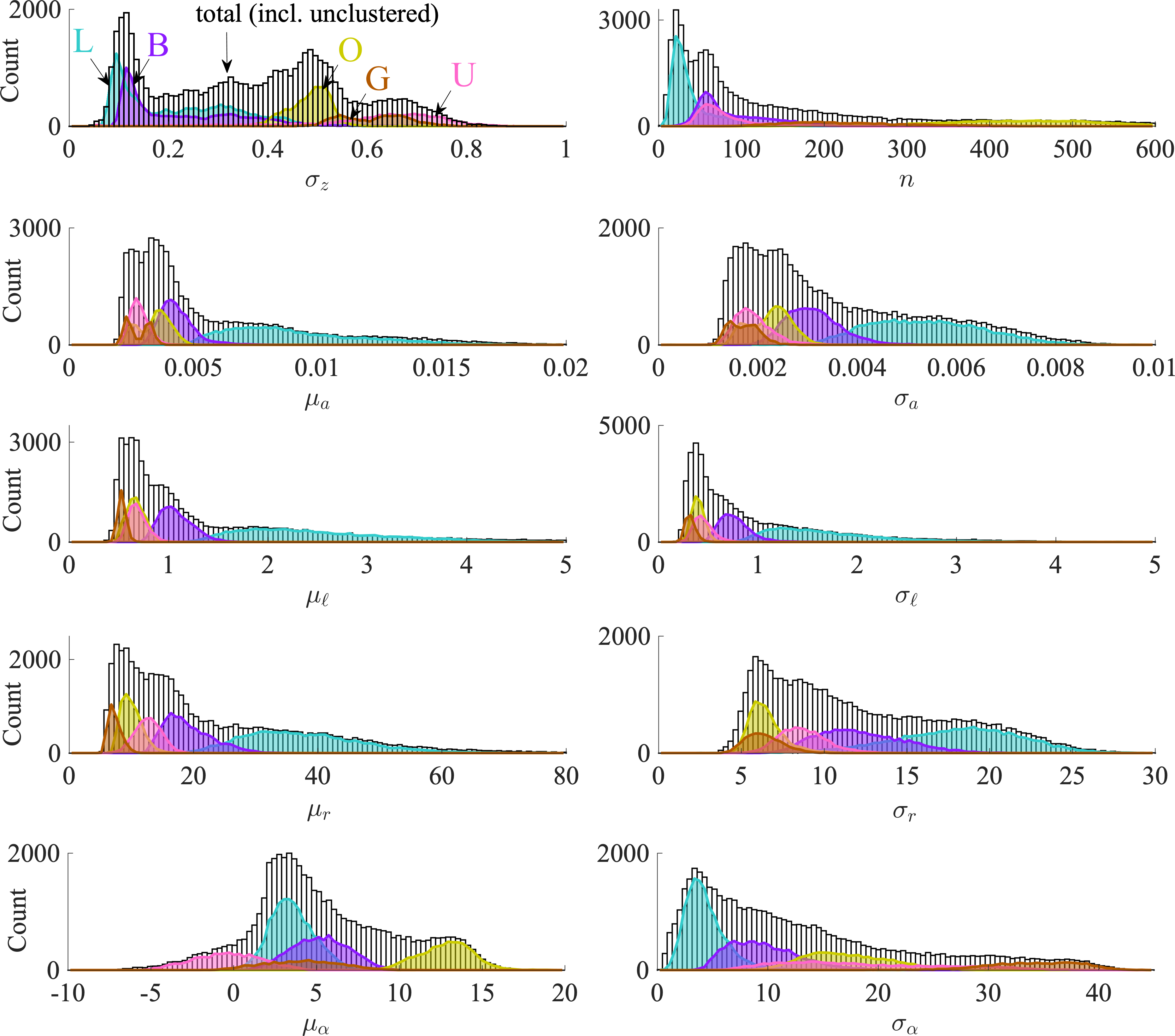}
\caption{(a) Density morphology properties (vector $\mathbf{m}$, see Eq.~\ref{eq:x}) of cluster centroids (black squares, Fig.~\ref{fig:clusters_PCA}), expressed as normalized deviations from the total distributions. (b) Histograms of the distributions and the contribution of each cluster in physical, non-dimensional space $\mathcal{M}$. }
\label{fig:histograms}
\end{figure}

Fig.~\ref{fig:histograms}a illustrates the values of the original ten \textit{normalised} morphology statistics  corresponding to the cluster centroids plotted in Fig.~\ref{fig:clusters_PCA}.
For each cluster, white denotes a property equal to the average of the total distribution of 50155 points, whereas blue and red colours denote properties below and above the average, respectively.  Fig.~\ref{fig:histograms}b shows histograms of the underlying distribution of \textit{physical (non-normalised)} morphology statistics $\mathbf{m} \in \mathcal{M}$ (see Eq.~\ref{eq:x}). White bars show the total distribution and colours show the contribution of each cluster. From a combined analysis of Fig.~\ref{fig:histograms}a-b we deduce the following  typical descriptions of each cluster. These descriptions will be illustrated with representative shadowgraph images from each cluster in Figure \ref{fig:clusters_edges}  (discussed in  \S~\ref{sec:edgesShadowgraphs}).

\begin{itemize}
    \item[L:] The density interfaces are typically concentrated in a single set (i.e. a unimodal distribution of edges having low $\sigma_z$), are scarce (low $n\approx 10-30$), but have a relatively large area (high $\mu_a$), are long (high $\mu_\ell\approx 1.5-3$), slender (high $\mu_r\approx 30-45$), and flat (low tilt $\mu_\alpha\approx 0-6^\circ$). They show significant spatial variability (in any instantaneous frame) in their area, length, aspect ratio (high $\sigma_a,\sigma_\ell,\sigma_r$), but not in their tilt (low $\sigma_\alpha$). These properties suggest relatively stable and laminar-like flow snapshots, as will be illustrated by a typical image in Fig.~\ref{fig:clusters_edges}.
    \item[B:]  The density interfaces have the most average properties among all clusters: average vertical spread $\sigma_z$, number ($n\approx 30-150$), length,  aspect ratio, and tilt, both in mean values and spatial variability (e.g. mean $\mu_\alpha\approx 2-10^\circ$ and standard deviation $\sigma_\alpha \approx 5-15^\circ$). These properties suggest less stable, intermediate snapshots.
    \item[O:] The density interfaces are fairly spread out in $z$  ($\sigma_z\approx 0.4-0.6$), are by far the most numerous ($n\approx 300-600$) and the most tilted ($\mu_\alpha\approx 9-16^\circ$, $\sigma_\alpha\approx 10-25^\circ$), but they are short ($\mu_\ell \approx0.5-1$) and thick ($\mu_r\approx 5-15$). These properties suggest very unstable snapshots which feature a large number of distinct density interfaces. 
    \item[G:] The density interfaces are greatly spread out in $z$ ($\sigma_z\approx 0.5-0.8$), and have strong variability in tilt ($\sigma_\alpha\approx 30-40^\circ$), but they are  the smallest, shortest ($\mu_\ell \approx 0.5$) and thickest ($\mu_r\approx 5-10$). They are about average in number ($n\approx 100-300$) and in mean tilt ($\mu_\alpha\approx 0-10^\circ$). These properties suggest turbulent mixing across a thicker layer than in cluster O but with fewer and more stable detectable interfaces, perhaps due to weaker density gradients.
    \item[U:] The density interfaces resemble those in cluster G, but are much  less numerous ($n\approx 40-120$) and flatter, with a mean tilt that is  -- uniquely -- frequently negative ($\mu_\alpha\approx -5^\circ$ to $5^\circ$) and very variable ($\sigma_\alpha \approx 7-30^\circ$). These properties suggest fewer distinct density interfaces, perhaps comprising primarily the strongest (and therefore most stable) density gradients on either side of the mixing layer, together with a few weaker and more three-dimensional, small-scale gradients within it. Once integrated across the entire spanwise direction in Eq. \eqref{eq:sg}, they significantly blur the mixing layer, resulting in few detected edges.  
\end{itemize}
The level of detail in which each cluster may be interpreted in terms of the morphological properties of its density interfaces is significantly more informative  than the prior qualitative descriptions of distinct flow regimes chosen by the human eye. These quantitative features appear to suggest increasing levels of turbulence from cluster L to U, perhaps most clearly shown by the histogram of $\sigma_z$ showing the increasing spread of density interfaces along $z$ from L to U. In the next section we confirm and illustrate these findings using the original shadowgraph and edge images.

\subsection{In terms of edges and shadowgraphs ($\mathcal{C}\rightarrow \mathcal{E}\rightarrow \mathcal{S}$)} \label{sec:edgesShadowgraphs}

\begin{figure}
\vspace{-0.5cm}
\centering
\includegraphics[width=\linewidth]{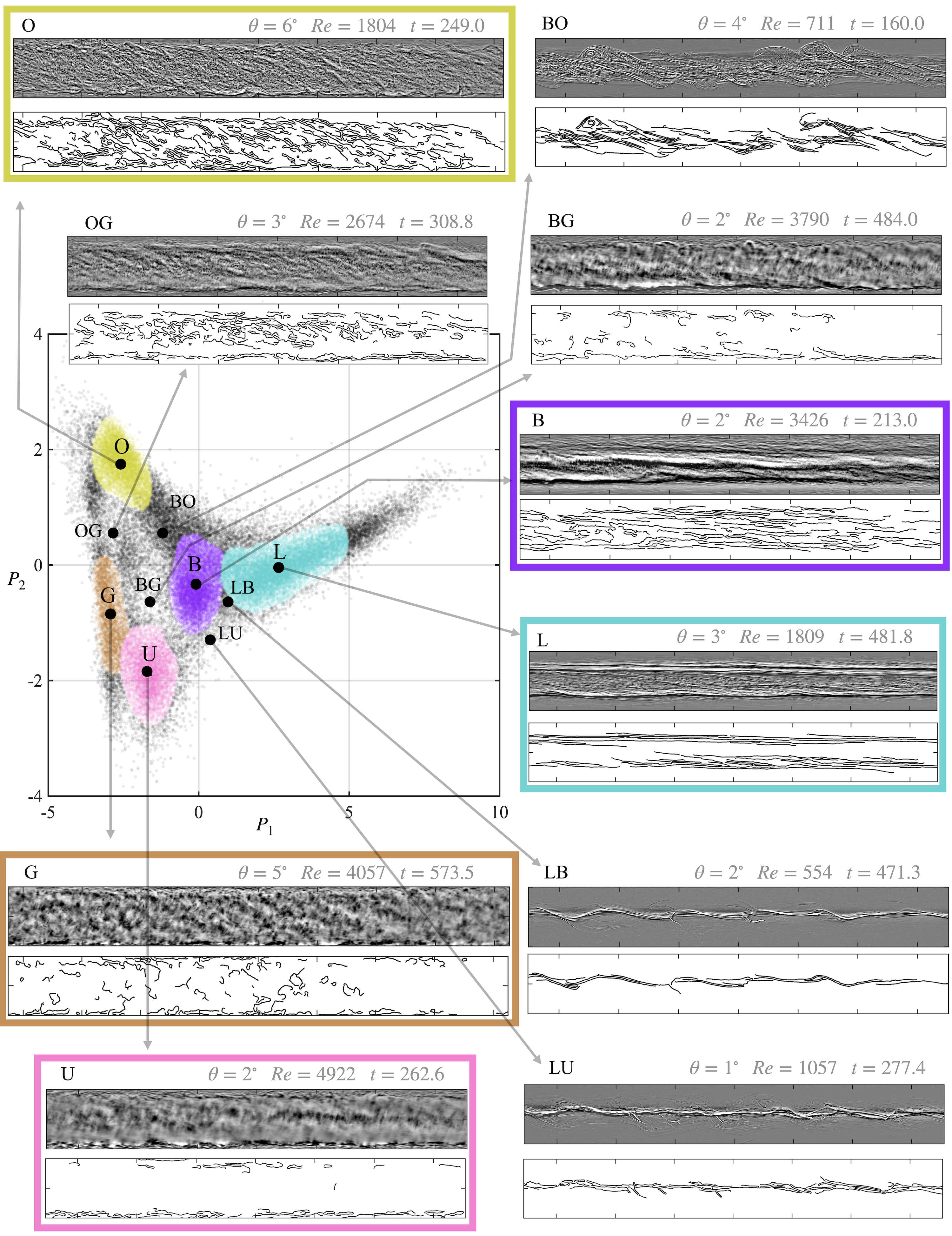}
\caption{Shadowgraph and corresponding edge images for the centroid of each cluster (coloured) and additional examples of frames between clusters. The input parameters $(Re,\theta)$ and time $t$ are indicated in the top-right corner of each frame.  Frames are scaled to have equal height, resulting in a range of lengths.} 
\label{fig:clusters_edges}
\end{figure}

Fig.~\ref{fig:clusters_edges} shows examples of shadowgraphs $\in \mathcal{S}$ and edges $\in \mathcal{E}$, consisting of the frames nearest to each of the five cluster centroids (labelled L to U in Fig.~\ref{fig:clusters_PCA}) and of five additional unclustered frames chosen to span the sparser, intervening unclustered regions (LB, BO, OG, BG, LU). 

Frame L illustrates the morphological description of the L cluster given in the previous section. However, it does not intuitively qualify as the most laminar flow. Its parameters ($\theta=3^\circ$ and $Re=1809$)  place it instead on the upper end of the intermittent regime (see Fig.~\ref{fig:setup}b). This frame thus captures a relaminarisation phase, as the shadowgraph still exhibits small-scale structure from a past turbulent phase, which is (only partially) mirrored in the edge image by the stacking of multiple flat and stable density interfaces. This frame represents what we may call \textit{laminarizing turbulence}. As a reference point, a stable laminar flow with a density profile having a single flat, sharp (pixel-thin) interface centred at $z=0$ would give $\mathbf{m} \approx [0, \ 1, \ 0.0056, \ 0, \ 16, \ 0, \ 3500, \ 0, \ 0, \ 0] \Rightarrow \mathbf{p} \approx [1287,  \ 248]$, i.e. very far off the top right vertex of the triangle in the $(P_1,P_2)$ plot.

Frames LB and LU intuitively qualify as the least turbulent -- despite their proximity to more turbulent clusters B and U  -- and illustrate the well-known Holmboe wave regime. The Holmboe instability is caused by the interaction of two vorticity waves on either side of a broad shear layer with an internal gravity wave on the sharper density interface at the centre of the shear layer \cite{carpenter_instability_2011}. This linear instability saturates nonlinearly at finite amplitude, giving a pair of counter-propagating modes with a distinctive cusped shape, which persist for arbitrarily long run times. At these amplitudes only minimal three-dimensional mixing takes place, hence these frames represent the `lower end' of what we may call \textit{Holmboe wave turbulence}. 

Frame B illustrates what we may call \textit{braided turbulence} owing to its pair of stacked, flat, central `braids' with a strong density curvature evidenced by the strong contrast in shadowgraph intensity (black and white shades). These two strong interfaces define an essentially three-layer density profile. Weaker contrasts (shades of gray) reveal a number of weaker interfaces, which do not extend all the way to the duct walls. 

Frame O illustrates \textit{overturning turbulence} owing to numerous short, unstable interfaces spanning most of the  duct height. Most `wisps' correspond to weak density curvature and contrasts in the shadowgraphs, which sets it apart from braided turbulence. Frame BO illustrates a transient roll up and mixing of two main braid-like interfaces (as in B) creating more tilted wisps (as in O).

Frames G and U illustrate what we may call \textit{granular turbulence} and \textit{unstructured turbulence}, respectively. Both have strong three-dimensional density curvature spanning the duct height. We recall from the reachability plot (Fig.~\ref{fig:reach}) and the $(P_1,P_2)$ plot (Fig.~\ref{fig:clusters_PCA}) that clusters G and U are adjacent and might in fact be nearly considered to be one larger single cluster, as these frames confirm. However, frame G has slightly stronger contrast and hence a larger number of detected edges, especially at mid-depth, some of which are nearly circular and give an overall granular appearance. The interfacial turbulence in frame U has much less structure, presumably because of higher three-dimensionality and dynamically-active lengthscales. The edges in U tend to be flatter than in G and located almost exclusively near the top and bottom walls, where the thick intermediate layer of mixed fluid meets boundary layers of relatively laminar and unmixed fluid.  The eye appears to detect in U a pair of two turbulent braids in lighter shades of gray sandwiching a slightly darker region, but this large scale pattern is too weak to be detected by the edges algorithm.

Frame OG illustrates an intermediate stage between overturning and  fully granular turbulence, featuring a mix of both types. We note that the set of input parameters $(Re,\theta)$ would not allow us to classify, a priori, the turbulence in OG as intermediate between O and G. Similarly, frame BG illustrates how braids (B) grow thicker and more blurred, as an intermediate stage before granularity (G). A frame BU halfway between clusters B and U would likely look similar. 

The data-driven discovery of these different types of turbulence constitute the first key finding of this paper.  Future improvements may consider classifying, within each frame, various ``dynamically distinct regions'' in the spirit of \cite{portwood_robust_2016} (who used the density gradient field). Unsupervised neural networks have also been used to detect the laminar/turbulent boundary in transitional boundary layers \cite{narasimhan_large_2021} (using the velocity field), while others subdivided the domain  and applied clustering to identify ``the regions containing streaks, turbulent spots [...] and developed turbulence'' \citep{foroozan_unsupervised_2021}.

\subsection{In terms of human classified regimes ($\mathcal{C}  \rightarrow \mathcal{H}$)}

\begin{figure}
\centering
\includegraphics[width=0.85\linewidth]{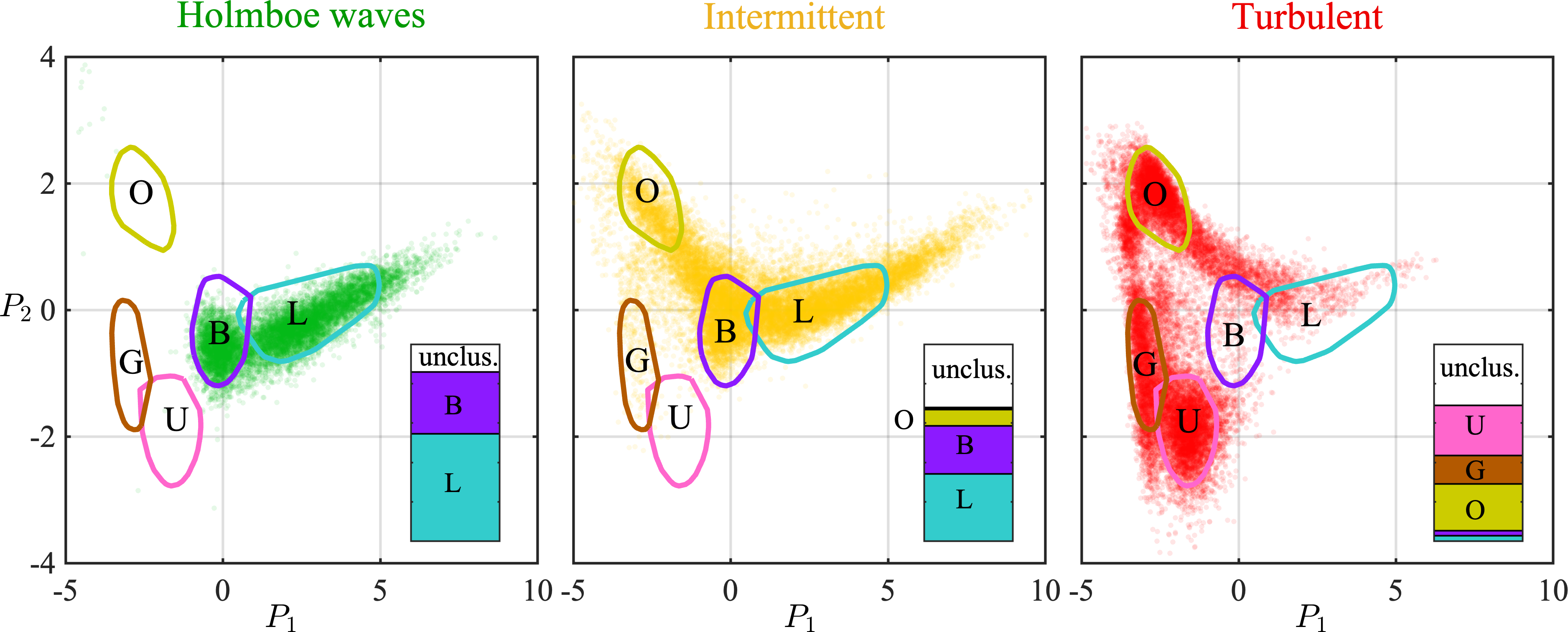}
\caption{Comparison between the location of clusters in principal component space  (approximated by their convex hull perimeter) and the three human-classified regimes (semi-transparent coloured circles): Holmboe wave (left), Intermittent (center) and Turbulent (right). Insets denote the fraction of frames in each cluster.}
\label{fig:regimes-clusters}
\end{figure}

Fig.~\ref{fig:regimes-clusters} shows the overlap in the two-dimensional principal component space $\mathcal{P}$ between the human-classified regimes H, I, T $\in \mathcal{H}$ (see $\S$ \ref{sec:human-class})  and the clusters L-U $\in \mathcal{C}$ obtained using our data-driven technique. All frames belonging to movies labelled `Holmboe' are coloured in green (left panel), and similarly for movies labelled `Intermittent' (middle panel, in yellow) and `Turbulent' (right panel, in red). The insets in each panel show the fraction of frames belonging to each cluster.

The Holmboe (H) regime spans only clusters L and B, with a few unclustered frames located either around the top right vertex of the triangle towards stable laminar flow or outside the lower boundary of L and B (see frames LB and LE in Fig.~\ref{fig:clusters_edges}). This means that waves and turbulence that are distinctively of Holmboe type display a range of density interfaces akin either to a slightly perturbed laminar flow, laminarizing turbulence (in L), or braided turbulence (in B). 

The intermittent (I) regime spans clusters L and B in approximately equal measure, although slightly more towards their upper boundary (contrary to the H regime), as well as, to a smaller extent, cluster O. It also contains a much larger proportion of unclustered frames, which are located around the top right vertex and between B and O. This means that flows classified as intermittently turbulent display a smooth continuum of most of the features we would expect: stable laminar flow, as well as overturning, braided, and laminarizing turbulence. 

The turbulent (T) regime spans clusters O, G, U in approximately equal measure, with the fourth quarter of frames being unclustered. This means that flows classified as fully turbulent feature more overturning  than any other flows, and that they are unique in displaying granular and unstructured turbulence. Moreover, the T regime has the widest distribution of frames across $\mathcal{P}$, covering all clusters to some extent as well as most of the unclustered regions. This suggests that the variability  (in time and across $\mathcal{I}$) of density interfaces is richer in the T than in the I regime. 

\subsection{In terms of input parameters ($\mathcal{C}\rightarrow \mathcal{I}$)}

\begin{figure}
\centering
\includegraphics[width=0.98\linewidth]{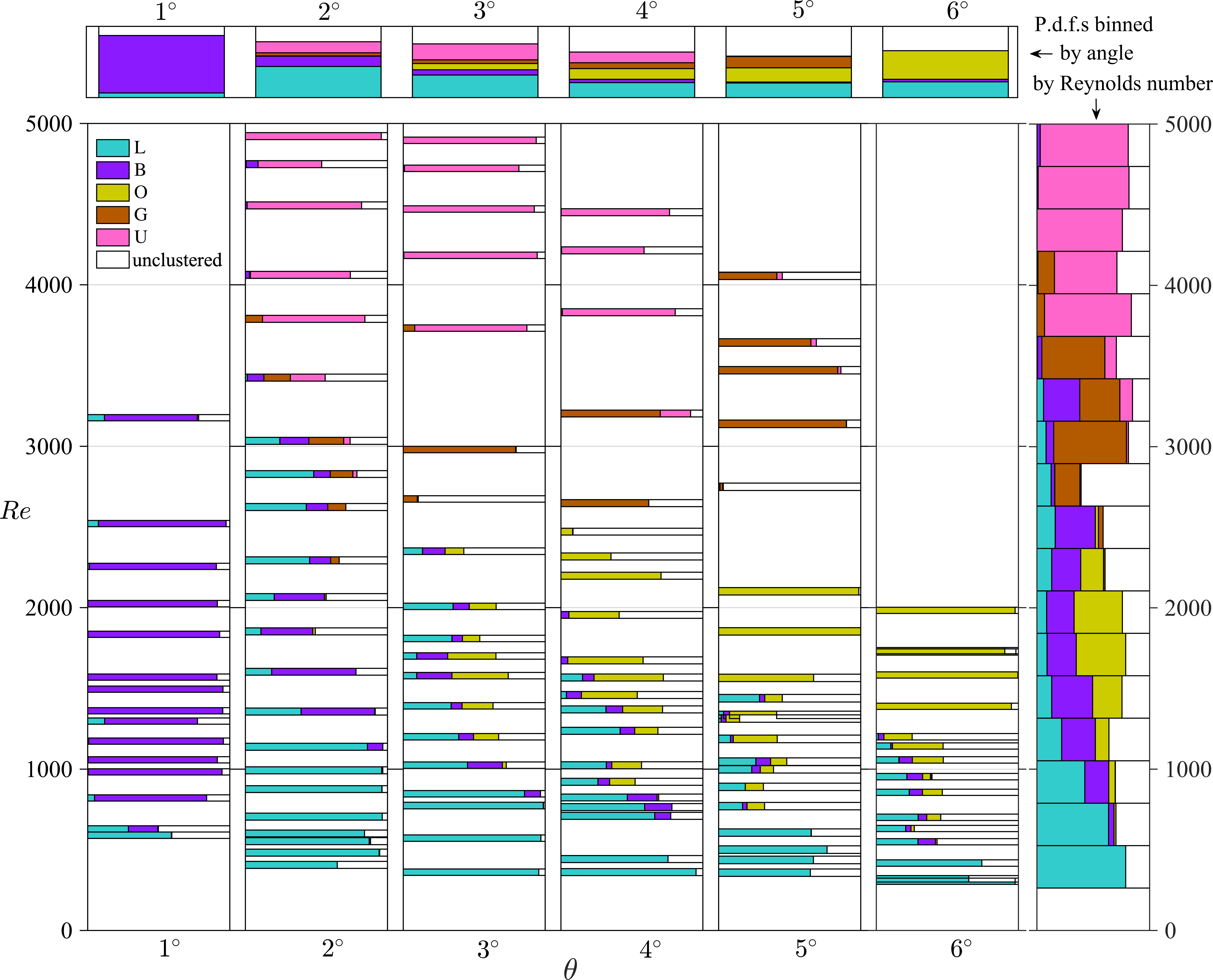}
\caption{Ratio of frames (time spent) in each cluster for each of the 113 experiments organized in parameter space $(Re,\theta)$. Some overlap occurs between bars at neighboring $Re$ values.  The top row and right column show overall p.d.f.s. The full time series of 21 representative experiments are shown in Fig.~\ref{fig:timeseries_PCA}.}
\label{fig:Re-theta}
\end{figure}

We now study the distribution of time spent in each cluster in Fig.~\ref{fig:Re-theta}, shown by a horizontal colored bar for each of the 113 experiments at the point $(Re,\theta)$ at which they were run. Complementary p.d.f.s corresponding to the entire data binned in $Re$ and $\theta$ are shown in the top and right columns of the plot, respectively.

Moving from left to right (increasing $\theta$) at intermediate $Re \approx 1500-3000$, we find almost exclusively braided turbulence (B) at $\theta=1^\circ$, which gradually gives way to more -- and eventually exclusively -- overturning turbulence (O) at $\theta=6^\circ$. Moving from the bottom to top (increasing $Re$) at intermediate $\theta \approx 2-5^\circ$, we find a gradual decrease in laminarizing turbulence (L), followed first by an increase in braided turbulence (B) at $Re\approx 500$ and then in overturning turbulence (O) at $Re\approx 1000$ (as well as intermediate,  unclassified types). We then find a decrease in unclassified turbulence and an increase in fully three-dimensional turbulence, first in cluster G at $Re\approx 2500$, and eventually in cluster U at $Re\gtrsim 3500$. The binned p.d.f.s highlight that this change of type of turbulence across $\mathcal{I}$ occurs through gradual shifts in the  time spent in the respective clusters, as $Re$ and $\theta$ are varied. 

This automated, quantitative characterization of the parameter space of high-$Re$ stratified turbulence constitutes the second key finding of this paper. Future improvements should note that our results have limited precision since the experimental dataset does not uniformly sample the entire $(Re, \theta)$ plane with fine resolution.  Further, although our time series are relatively long (on average $444$ A.T.U.), even longer time series would help improve convergence towards the full underlying dynamical picture.


\subsection{Temporal intermittency and transitions} \label{sec:intermittency}

\begin{figure}
\centering
\includegraphics[width=0.86\linewidth]{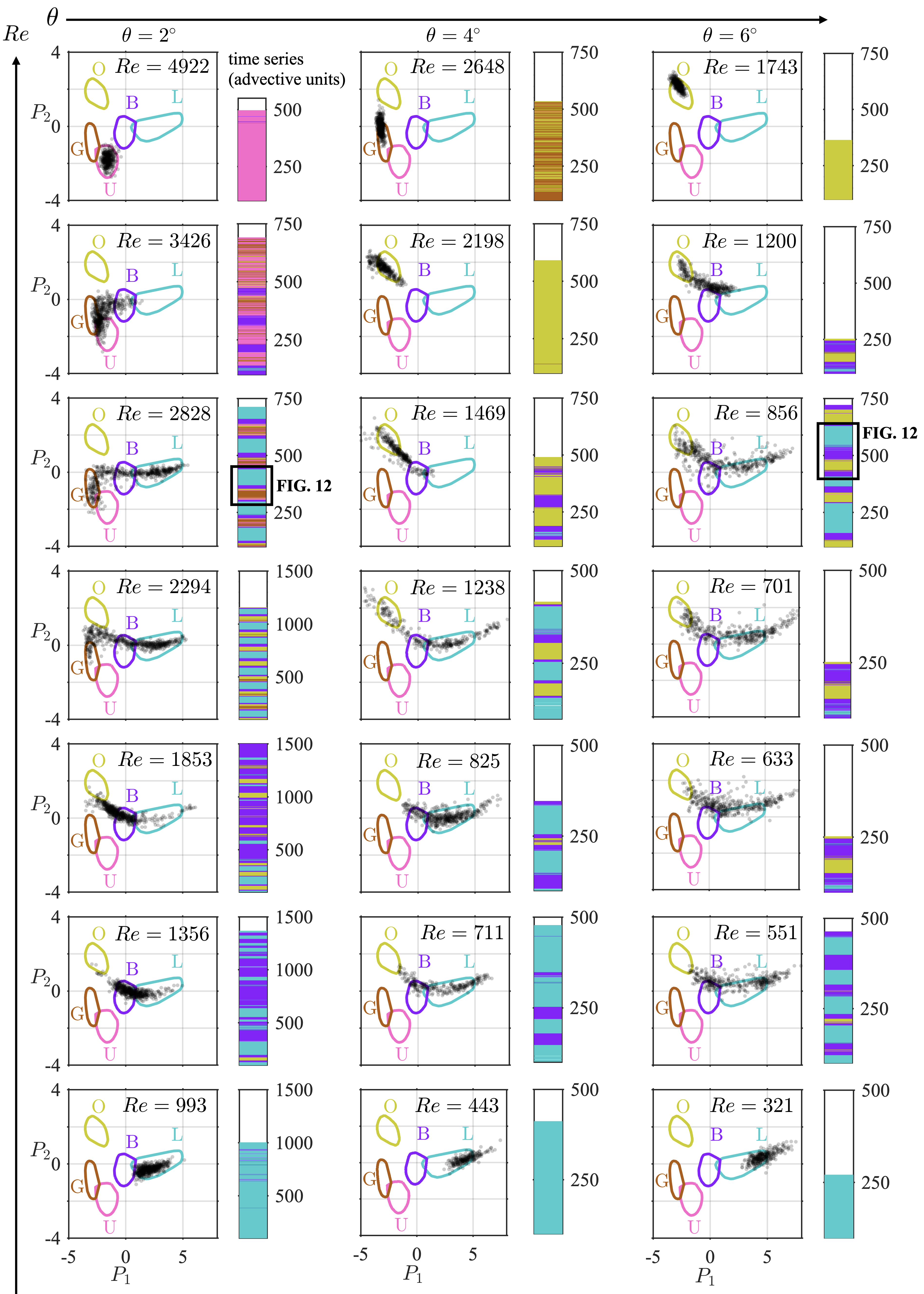}
\caption{Temporal dynamics in experiments at increasing $Re$ (from bottom to top) at three tilt angles: $\theta=2^\circ$ (left column),  $\theta=4^\circ$ (middle) and  $\theta=6^\circ$ (right). Individual frames are marked by transluscent black symbols and cluster boundaries (convex hulls) are denoted in color. Time series indicating which cluster the frames belong to, for $t\ge 100$, are shown by a vertical bar on the right. Unclustered frames are colored based on the nearest cluster. Note the involuntary variations in recording time between experiments. }
\label{fig:timeseries_PCA}
\end{figure}

Fig.~\ref{fig:timeseries_PCA} examines the temporal dynamics of 21 experiments selected across the input space $\mathcal{I}$ (see Fig.~\ref{fig:Re-theta}), with increasing $Re$ from bottom to top, and increasing $\theta$ from left to right. Each panel shows the frame-by-frame trajectories in phase space $\mathcal{P}=(P_1,P_2)$ using translucent black symbols and each vertical bar shows the corresponding time series for which cluster each frame belongs to (for clarity, unclustered frames are colored based on the cluster having the nearest centroid). Each row shows a set of three experiments with approximately matched $Re\, \theta$, decreasing from $\approx 1\times 10^4$ (top row) to $\approx 2\times10^3$ (bottom row). Previous SID theory, experiments \cite{lefauve_regime_2019,lefauve_experimental2_2022} and direct numerical simulations \cite{zhu_stratified_2023}  showed that the product $Re \, \theta$ is proportional to the dynamic range of stratified turbulence, i.e. the buoyancy Reynolds number $Re_b=(L_O/L_K)^{4/3}$, measuring the separation between the Ozmidov $L_O=(\epsilon/N^3)^{1/2}$ and Kolmogorov $L_K=(\nu^3/\epsilon)^{1/4}$ turbulent lengthscales, where $\epsilon$ is the averaged turbulent kinetic energy dissipation and $N$ is the averaged buoyancy frequency \cite[Sec. 5.1]{lefauve_experimental2_2022}.

Tracking the most dissipative turbulence along the top row of Fig.~\ref{fig:timeseries_PCA} (highest $Re\,\theta$), we find different types of turbulence, each of which is tightly grouped in phase space. At the lower angle $\theta=2^\circ$, turbulence is exclusively unstructured, staying within or very near cluster U. At the intermediate $\theta=4^\circ$, it shifts to being granular (primarily in G) with brief excursions to the sparser (unclustered) space towards O. At the higher $\theta=6^\circ$, it is exclusively of overturning type (in or near O). At slightly lower values of $Re\,\theta$ (second row), the trajectories are less tightly grouped in phase space and intermittency appears. At $\theta=2^\circ$ and $Re=3426$, turbulence now cycles between U, G and B.  At $\theta=4^\circ$ and $Re=2198$, turbulence shifted to O, but without clear intermittency.  At $\theta=6^\circ$ and $Re=1200$, turbulence is intermittent between O, B and L (despite the short time series for this dataset).

At slightly lower values of $Re$ again (third row), the trajectories are even more spread out and intermittency is now generic at all angles. At $\theta=2^\circ$ and $Re=2828$, the  grey data cloud now covers G, B and L, with a distinctive `upward bend' between  G and B (avoiding the O cluster). The time series cycles quasi-periodically between them with period $T \approx 120-140$ advective units (further details are given in the companion Letter \cite{lefauve_routes_2023}, including the residence time in each cluster). Importantly, the transitions from L to G always pass through B, just like the transitions from G to L turbulence; moreover both transitions follow identical trajectories in this two-dimensional projection of phase space. At $\theta=4^\circ$ and $\theta=6^\circ$, the grey data cloud assumes a fundamentally different shape, between the O, B, and L clusters, with a distinctive `downward bend'. The time series are again quasi-periodic, with period $T\approx 120-220$.  At $\theta=6^\circ$ and $Re=856$, the relaminarizations are more complete, and the transitions to and from O pass through B with identical trajectories, just like at $\theta=2^\circ$. Moving down in $Re$ again (fourth to seventh row), we find that these two types of intermittency persist for a wide range of $Re$, but that the most turbulent phases (in G at $\theta=2^\circ$ and in O at $\theta=4^\circ, 6^\circ$) gradually become shorter and are replaced by longer phases in B (fourth to sixth row), and eventually in L (seventh row), which here corresponds in fact to Holmboe waves. 

\begin{figure}
\centering
\includegraphics[width=\linewidth]{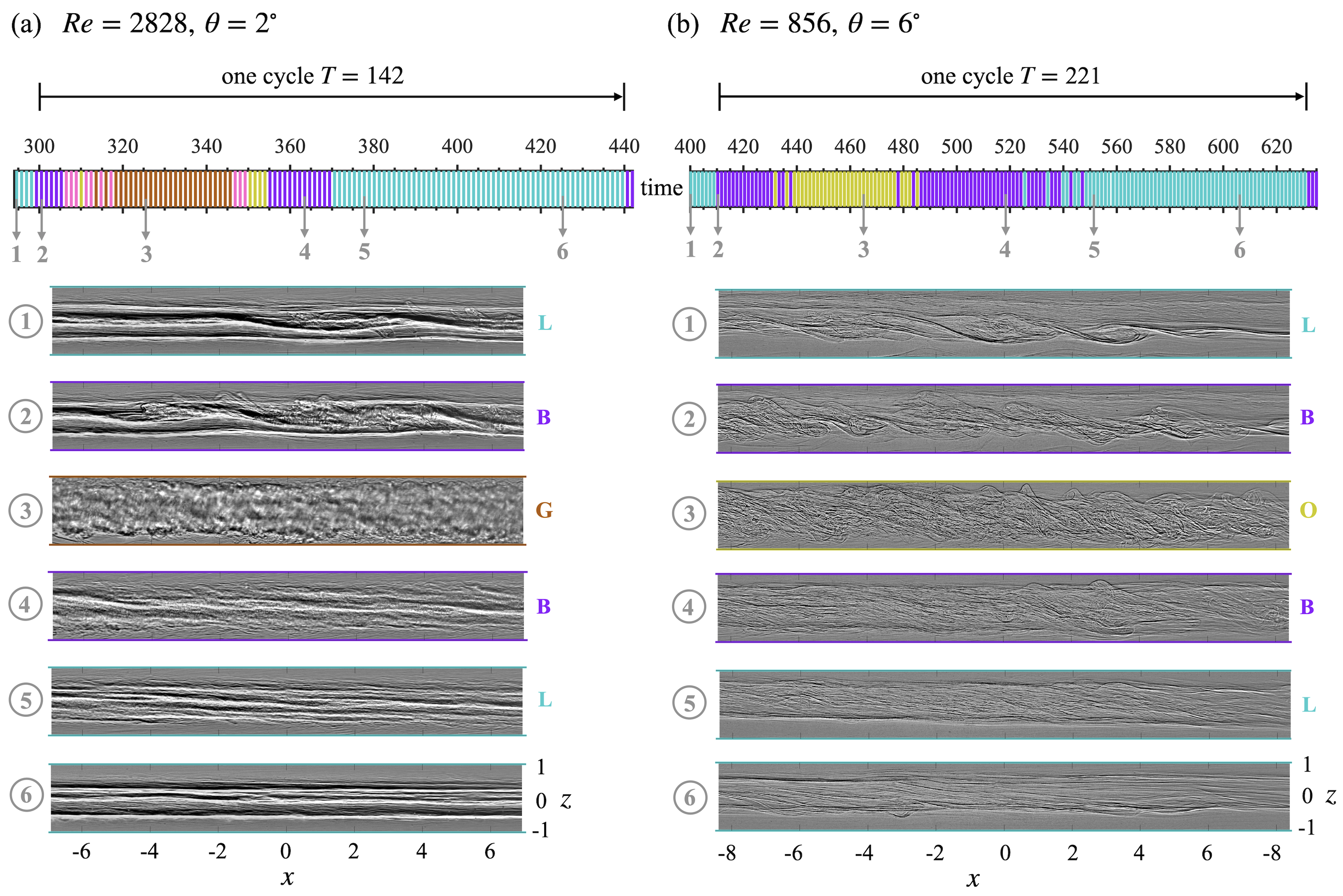}
\caption{Detailed transitional states present in the two distinct intermittent cycles identified in Fig.~\ref{fig:timeseries_PCA} (left and right column, third row). We contrast the G-B-L dynamics (left column) to the O-B-L dynamics (right) during a single cycle. Time series are provided in the top bar of each column (colored by nearest cluster) while frames 1-6 illustrate shadowgraphs at six representative times. }
\label{fig:timeseries_transitions}
\end{figure}

The two distinct quasi-periodic dynamics at intermediate $Re$ (G-B-L vs O-B-L) reveal two fundamentally different routes to turbulence in SID, and constitute the third key finding of this paper. SID intermittency appears to organizes around at least two inherently different `slow manifolds' in different regions of $(Re,\theta)$. Future work is still needed to characterize these slow manifolds and their corresponding `orthogonal' fast manifolds, in order to understand the specific flow structures responsible for the laminar/turbulent transition in different regions of $(Re,\theta)$.

Fig.~\ref{fig:timeseries_transitions} makes a first step in this direction by contrasting these two distinct transitional phases by zooming in on two quintessential time series of Fig.~\ref{fig:timeseries_PCA} (black boxes, first and third columns of third row) and showing six representative shadowgraph frames spanning a single cycle. Just before the start of a cycle  (frames numbered `1', Fig.~\ref{fig:timeseries_transitions}), signs of a growing instability are weak enough that the density interfaces are still classified as laminarizing turbulence (L). At the start of a cycle (frames 2), the instabilities have grown in amplitude and are now classified as braided turbulence (B). However, we find significant differences between the left and right frames, representing distinct routes to turbulence. The left frame shows turbulence arising first from localized small-scale structures on the upper density interface, while the right frame shows turbulence arising from a more extensive roll-up.  These two transitions add complexity to our understanding of stratified turbulence and to the classical paradigm of a Kelvin-Helmholtz breaking billow, featuring a `hot' growing phase, a `Goldilocks' energetic phase, and a `cold' fossilization decaying phase \cite{mashayek_goldilocks_2021,smith_turbulence_2021}. After the active turbulence phase (frames 3), the stabilising phase significantly differs between the two experiments, from the first stage in B (frames 4) to the last stage in L (frames 5 and 6). Turbulence and mixing subside, leaving a three-layer stratification with a partially-mixed intermediate layer having a complex, temporally-evolving structure. These particular time series (top bars) also highlight the general feature that the excursions in B during the relaminarization phase G/O $\rightarrow$ B $\rightarrow$ L are consistently longer than during the unstable transitional phase L $\rightarrow$ B $\rightarrow$ G/O.

The above finding that apparently distinct transitions pass through the same cluster (either B or L, see Fig.~\ref{fig:timeseries_transitions}) suggests that consideration of a higher-dimensional phase space (greater than the two-dimensional space $\mathcal{P}=(P_1,P_2)$ considered here) may yield a deeper understanding of the underlying higher-dimensional intermittent dynamics.
Future work considering a higher-dimensional space of 
principal components $\mathcal{P}=(P_1,P_2,P_3,\ldots)$ may thus be able to resolve the bursting and relaxation dynamics in the fast manifold \cite[\S~6.7.2]{schmid_data-driven_2021} that may be orthogonal to the first two principal vectors.  Candidates for this analysis are cluster-based Markov chain models \cite{kaiser_cluster-based_2014,foroozan_unsupervised_2021}, network models \cite{li_cluster-based_2021}, and the Perron-Frobenius (transfer) operator approximated with Ulam's method \cite{junge_discretization_2009}. The latter has the potential to identify the density interface structures ``that have a higher probability of detaching from the slow manifold and use these structures as precursors (or predictors) of impending violent events'' \cite{schmid_data-driven_2021}.

\section{Conclusions} \label{sec:ccl}

We have developed a method for data reduction and automatic image classification and applied it to 50155 individual frames sampled from 113 shadowgraph movies from the Stratified Inclined Duct (SID) experiment, sustaining sheared stratified turbulence at high Reynolds and Prandtl numbers. These movies provide, over hundreds of advective time units, a spanwise-integrated view of the curvature of the density field (caused by a spatially-varying salinity) allowing us to identify various turbulent states based on a clustering of the morphology of density interfaces embedded within the flow. The three key results in Sec.~\ref{sec:physical_interp} can be summarized as follows.

\textit{Instantaneous classification of turbulence} -- A physical interpretation of the identified clusters in Fig.~\ref{fig:clusters_edges} revealed five distinct types of stratified turbulence and mixing: laminarizing (L), braided (B), overturning (O), granular (G) and unstructured (U), as well as intermediate types. The strength of our automated classification approach is that it is objective, quantitative, sensitive to fine details of the flow, and can be readily applied to a vast number of instantaneous frames. Its results differ from and complement the prior human classification of movies into the Holmboe wave, intermittently-turbulent  and turbulent flow regimes. Our clustering reveals that flows belonging to the same regime generally have a different temporal `mix' of types of turbulence as the  parameters ($Re,\theta$) are varied within the regime. 
Our data-driven approach can also be easily generalized and adapted to classify stratified turbulence in other insightful experiments, such as the stratified Taylor-Couette flow \cite{caton_primary_1999,oglethorpe_spontaneous_2013,leclercq_using_2016,ibanez_observations_2016,park_competition_2018,petrolo_buoyancy_2020,meletti_experiments_2021}.


\textit{Dynamical map of parameter space} -- The fractions of time spent in each cluster in Fig.~\ref{fig:Re-theta} followed gradual variations across the input space $(Re,\theta)$. 
Simply speaking, by increasing $Re$, laminarizing turbulence gradually gives way to more braided turbulence, then overturning turbulence, and eventually granular and unstructured turbulence. By increasing $\theta$, braided, near-horizontal turbulence gradually gives way to overturning turbulence. This finding confirms the hypothesis in \cite[\S~6.4]{lefauve_experimental1_2022} (who used a dataset of simultaneous, three-dimensional velocity and density measurements in 16 experiments) that high-$Re$/low-$\theta$ turbulence has more extreme enstrophy events and that low-$Re$/high-$\theta$ turbulence has more density overturnings. Recalling that the product $Re\, \theta$ controls the rate of dissipation of turbulent kinetic energy, and that high-$Re$ flows have a wider spectral inertial subrange, it follows that high-$Re$ flows dissipate comparatively more at small scales (increasing extreme enstrophy) while high-$\theta$ flows dissipate comparatively more at large scales (increasing overturning). These differences in overturns statistics likely affect the energetics and efficiency of mixing \cite{mashayek_goldilocks_2021,mashayek_physical-statistical_2022}.

\textit{Distinct temporal routes to  turbulence} --- The phase-space trajectories of Fig.~\ref{fig:timeseries_PCA} revealed different dynamical behaviors with decreasing energy dissipation and dynamic range $Re\,\theta$. The most dissipative turbulence remains localized in cluster U, G or O (at low, medium and high $\theta$, respectively) for hundreds of advective time units. By contrast, in less dissipative turbulence, temporal intermittency gradually appears as the trajectories cycle between clusters G-B-L (at low $\theta$) and O-B-L (at high $\theta$), generically with remarkable quasi-periodicity. Shadowgraph snapshots at selected times during the cycles in Fig.~\ref{fig:timeseries_transitions} illustrated these two different transition pathways to and from stratified turbulence, i.e. laminarizing turbulence `curving' in phase space either towards granular or overturning turbulence and back. We hypothesised the existence of a low-dimensional slow manifold composed of L and O/G and of a faster, higher-dimensional manifold currently projected onto B. This paves the way for a future more accurate identification of the structures responsible for the quasi-periodic cycles of instability, turbulence, and relaminarization.

\textit{Outlook: reduced-order modeling} --- Returning to the approach introduced in expression (1) of the Introduction, we conclude that this paper has advanced the characterization of coherent structures and their link to input parameters (step 1). The way in which the different density interface morphologies identified here, the time spent in and between each cluster, and the intermittent flow history affect the useful output variables such as mixing (step 2) remains a fundamental challenge in the community \citep{caulfield_open_2020,caulfield_layering_2021}. This challenge -- reduced-order modeling -- remains to be tackled with full velocity and density datasets, now available in SID experiments \citep{lefauve_regime_2019} and direct numerical simulations \citep{zhu_stratified_2023}. Further data-driven techniques could also learn the mapping between cluster dynamics and measures of mixing to yield predictions in unseen datasets. For example, \cite{salehipour_deep_2019} used a deep convolutional neural network to learn the relation between a low-dimensional representation of turbulence (vertical profiles of buoyancy frequency and turbulent kinetic energy dissipation) and mixing efficiency, outperforming standard parameterizations.

\vspace{0.5cm}


\textbf{Data} --- All data will be made available.

\acknowledgements{We thank Xianyang Jiang, Gaopan Kong and the technicians of the G. K. Batchelor Laboratory for their help with the experiments. We are also grateful to Paul Linden and  Stuart Dalziel for their support  and to Colm-cille Caulfield for insightful discussions on the implications of this work.  The experimental facility was funded by the ERC grant  ‘Stratified Turbulence And Mixing Processes’ (STAMP, No 742480). A.L. acknowledges funding from a Leverhulme Early Career Fellowship and a NERC Independent Research Fellowship (NE/W008971/1). } 

\appendix

\section{Methods (complementing  Sec.~\ref{sec:detection-morph-density-int})}


\subsection{Post-processing of shadowgraph data} \label{sec:appendix-data-processing}

The following four steps ensured that all 113 raw movies (taking a total of $\approx$ 2 TB storage) could be used efficiently for automated classification. First, each movie was cropped vertically precisely at $z=\pm 1$ to only keep the internal wall-to-wall flow, resulting in a typical frame resolution of 3400 $\times$  450 pixels ($\approx 1.5$ MPixels with 8-bit depth, i.e. 1.5 MB each). Second, the temporal mean signal (background pattern) was removed  $I'(x,z,t)=I(x,z,t)-\langle I\rangle_t(x,z)$, and each frame was rescaled by calculating the 5th percentile $I'_{5}$ and 95th percentile $I'_{95}$ of the distribution of $I'$ and setting those of the rescaled frame as 0 and 1, respectively, by $\tilde{I} = I'/(I'_{95}-I'_5)$. All pixels having $\tilde{I}<0$ (5 \% of the data) were set to 0, and all pixels having $\tilde{I}>1$ (5 \% of the data) were set to 1. This ensured that all frames had comparable brightness and dynamic range, and that density interfaces could be later extracted with identical edge detection parameters. Third, we discarded the first 100 A.T.U. of each movie (keeping only data for $t\ge100$) to conservatively remove any transients associated with initial gravity currents (reaching the ends of the duct $x=\pm 40$ at an estimated $t\approx 80$ due to their speed $\approx 0.5$). Fourth, we subsampled the time series by a factor of 10, to avoid excessively redundant temporal data. This yielded an average temporal resolution across all movies of $0.10\times 10 = 1.0$ A.T.U. (standard deviation 0.4). The final dataset of 50155 frames takes $\approx 75$ GB storage.

\subsection{Principal components analysis} \label{sec:appendix-pca}

The data matrix $\mathbf{M}$ is first normalized to transform the values of the 10 morphology statistics into standard scores (or `z-scores'), where 0 corresponds to the mean value across all frames, and $\pm \lambda$ to $\lambda$ standard deviations above and below the mean, thereby ensuring that all characteristics are weighted equally. The singular value decomposition (SVD) of the normalized data matrix is then $\tilde{\mathbf{M}}= \mathbf{U}\boldsymbol{\Sigma}\mathbf{V}^T$. Here $\boldsymbol{\Sigma}=\{ \sigma_m \}_{m=1,\ldots,10}$  is the diagonal matrix of positive singular values arranged in decreasing order, and $\mathbf{U}= \{ \mathbf{u}_m \}_{m=1,\ldots,10} \in \mathbb{R}^{50155\times10}$ and $\mathbf{V}= \{ \mathbf{v}_m \}_{m=1,\ldots,10} \in \mathbb{R}^{10\times10}$ are the real orthogonal matrices containing the left and right singular vectors, respectively.  The cumulative variance in Fig.~\ref{fig:correlation_PCA}b shows that $\Sigma(2)/\Sigma(10) =0.79$, noting that the total variance $\Sigma(10)$ is the sum of the 10 eigenvalues of the covariance matrix $\tilde{\mathbf{M}}^T\tilde{\mathbf{M}}$. This justifies the truncated approximation  $\tilde{\mathbf{M}}\approx \tilde{\mathbf{M}}_t =\mathbf{U}_t\boldsymbol{\Sigma}_t\mathbf{V}_t^T$ where we only retain the first two columns of  $\mathbf{U}_t,\mathbf{V}_t$  and  the first $2\times2$ block of $\boldsymbol{\Sigma}_t$. In other words, $\tilde{\mathbf{M}}_t = \sigma_1 \mathbf{u}_1 \mathbf{v}_1^T+\sigma_2 \mathbf{u}_2 \mathbf{v}_2^T$, which approximates each frame (row) $f=1,\ldots,50155$ as the sum $\approx \sigma_1 u_{f1}\mathbf{v}_1^T +\sigma_2 u_{f2}\mathbf{v}_2^T$ using the top two right transposed singular vectors $\mathbf{v}_1^T,\mathbf{v}_2^T \in \mathbb{R}^{1\times 10}$. Rewriting this as $p_1\mathbf{v}_1^T +p_2\mathbf{v}_2^T$ highlights that we effectively project each rank-two-approximated frame from the 10-dimensional space $\mathcal{M}$ to the two-dimensional space $\mathcal{P}$  spanned by the two normal unit vectors $\mathbf{v}_1^T,\mathbf{v}_2^T$, and obtain the vector of coordinates $\mathbf{p}=[p_1,p_2]=[\sigma_1 u_{f1},\sigma_2 u_{f2}]$. In matrix form, this mapping thus transformed $\tilde{\mathbf{M}}\in \mathbb{R}^{50155\times 10}$ to the new data matrix $\mathbf{P} = \tilde{\mathbf{M}}_t \mathbf{V}_t =\mathbf{U}_t\boldsymbol{\Sigma}_t \in \mathbb{R}^{50155\times 2}$ containing the 50155 two-dimensional row vectors $\mathbf{p}$. 


\subsection{OPTICS algorithm} \label{sec:appendix-optics}

OPTICS computes the pairwise distances between all points based on a metric called the reachability distance:
\begin{equation}
    d_\text{reach} (\mathbf{p}_i, \mathbf{p}_j ) = \max \, \big( \, d_\text{euclidean} (\mathbf{p}_i, \mathbf{p}_j), \ d_\text{core} (\mathbf{p}_i) , \ d_\text{core} (\mathbf{p}_j )  \, \big),
\end{equation}
where $d_\text{euclidean}$ denotes the standard Euclidean distance, and $d_\text{core}$ denotes the core distance, i.e the radius of the hypersphere around each point (in our case a two-dimensional vector $\mathbf{p}$) that encloses exactly minPts neighbors.
The key property of the reachability distance is that it penalises points in sparser regions having larger $d_\text{core}$ by increasing their perceived distance from points in denser regions. A small $d_\text{core}$ indicates that the point sits in a dense regions of the space $\mathcal{P} = (P_1,P_2)$. OPTICS thus requires one user-specified parameter `minPts', which can be interpreted as the minimum number of points required to form a cluster.  Extreme values of $50155$ (our total number of points) or 1 would yield a single cluster or a cluster per point, respectively. To find a useful intermediate value, we progressively decreased minPts, leading to distinct valleys (meaningful clusters) in the reachability plot (as shown in Fig.~\ref{fig:reach}), which are robust for $600\lesssim$ minPts $ \lesssim 1200$. Decreasing minPts below $600$ leads to a much greater number of clusters clearly dominated by noise. The results in this paper were  computed using minPts $=800$. 





%

\end{document}